\documentclass{amsart}
\date{\today}

\newcommand{\lsim}{{\underset{\sim}{<}}}
\newcommand{\gsim}{{\underset{\sim}{>}}}
\newcommand{\q}{\qquad}

\newcommand{\R}{\mathbb{R}}
\newcommand{\C}{\mathbb{C}}

\newcommand{\Con}{C_0^\infty}

\newcommand{\grad}{\nabla}

\newcommand{\sa}{{{\pmb{a}}}}

\newcommand{\AAA}{{\text{\aa}}}
\newcommand{\AAAA}{{\text{\bf{\aa}}}}
\newcommand{\B}{{\mathbf{B}}}
\newcommand{\al}{{\pmb{\alpha}}}
\newcommand{\dv}{{\mathbb{D}_{V}}}
\newcommand{\pw}{{\mathbb{P}_{W}}}
\newcommand{\dvo}{{\mathbb{D}_{V_0}}}
\newcommand{\rvo}{{\mathbb{R}_{V_0}}}

\newcommand{\bsig}{{\pmb{\sigma}}}

\newtheorem{Lemma}{Lemma}
\newtheorem{Theorem}{Theorem}
\newtheorem{Corollary}{Corollary}
\newtheorem{Proposition}{Proposition}
\begin{document}

\title[Pauli and Dirac operators with large magnetic
fields]{Semi-classical
asymptotics for the counting functions and Riesz means of Pauli and Dirac operators
with large magnetic fields}
\author[A.~A.~Balinsky]{A.~A. Balinsky}
\address{School of Mathematics\\
         Cardiff University\\
         23 Senghennydd Road\\
         Cardiff CF2 4YH\\
         UK}
\email{BalinskyA@cardiff.ac.uk}
\author[W.~D.~Evans]{W.~D. Evans}
\address{School of Mathematics\\
         Cardiff University\\
         23 Senghennydd Road\\
         Cardiff CF2 4YH\\
         UK}
\email{EvansWD@cardiff.ac.uk}
\author[R.~T.~Lewis]{Roger T. Lewis}
\address{Department of Mathematics\\
         University of Alabama at Birmingham\\
         Birmingham, AL 35294-1170\\
         USA}
\email{lewis@math.uab.edu}

\begin{abstract}
We study the asymptotic behavior, as Planck's constant $\hbar\to 0$, of
the number of discrete eigenvalues and the Riesz means of Pauli and Dirac
operators with a magnetic field $\mu\B(x)$ and an electric field. The
magnetic field strength $\mu$ is allowed to tend to infinity as $\hbar\to 0$.
Two main types of results are established: in the first  $\mu\hbar\le constant$
as $\hbar\to 0$, with magnetic fields of arbitrary direction; the
second results are uniform with respect to  $\mu\ge 0$ but the magnetic
fields have constant direction. The results on the Pauli operator
complement recent work of Sobolev.
\end{abstract}

\keywords{Semi-classical asymptotics, Pauli and Dirac operators, counting functions, Riesz means}

\maketitle

\section{Introduction}

The Dirac and Pauli operators $ \dv, \pw $, which are the objects of
study in this paper, are defined as follows :
\begin{eqnarray}
\dv \equiv \dv(\B) &:= &\al\cdot (\frac{\hbar}{i}\mathbf{\grad}-\mu\sa)+\beta
  +V\nonumber\\
&\equiv &\sum_{k=1}^3{\al}_k(-i\hbar {\partial}_k
-\mu{\sa}_k)+\beta+V \\
\pw \equiv \pw(\B) &:= &[\bsig\cdot(-i\hbar\grad -\mu\sa)]^2 + W
\nonumber\\
&\equiv & [\sum_{k=1}^3{\bsig}_k(-i\hbar {\partial}_k
-\mu{\sa}_k)]^2+W\nonumber\\
&= & H_0( \mathbf{B}) -\mu \hbar \bsig \cdot
\mathbf{B}+W,
\end{eqnarray}
where $ H_0( \mathbf{B})= (-i\hbar \mathbf{\grad}-\mu
\mathbf{a})^2 $ is the Schr\"{o}dinger operator with magnetic field $
\mathbf{B}$ and
\begin{itemize}
\item
 $\sa = (a_1,a_2,a_3)$ is a magnetic vector potential with
magnetic field $\B:= \grad\times \sa $;
\item
 $\bsig =(\sigma_1,\sigma_2,\sigma_3)$ is the triple of Pauli matrices
$$
\sigma_1 = \left(\begin{matrix} 0 & 1\\ 1 & 0
\end{matrix}\right),\ \sigma_2 = \left(\begin{matrix} 0 &-i\\ i &
0 \end{matrix}\right),\ \sigma_3 = \left(\begin{matrix} 1 & 0\\ 0
&-1 \end{matrix}\right);
$$
\item
 $ \al = (\alpha_1,\alpha_2,\alpha_3) $ where $\alpha_j$,
 $j=1,2,3,$ and $\beta$ are the Dirac matrices
$$
\alpha_k = \left(\begin{matrix} 0_2 & \sigma_k\\ \sigma_k & 0_2
                    \end{matrix}\right),\ k=1,2,3,\q
                    \beta = \left(\begin{matrix} I_2& 0_2 \\ 0_2 & -I_2
                    \end{matrix}\right),
$$
where $0_2,I_2 $ are the $2\times2 $ zero and unit matrices
respectively.
\end{itemize}

Given a domain $\Omega \subseteq \R^3 $, self-adjoint realizations
$\pw(\Omega) $ and $\dv(\Omega) $ are defined by Dirichlet
boundary conditions on $\Omega$, these being satisfied in the
usual weak sense (see \S 2). The Pauli operator $\pw(\Omega)
$ acts in $L^2(\Omega)\otimes\C^2 \equiv [L^2(\Omega)]^2 $ and the Dirac
operator $\dv(\Omega) $ in $ L^2(\Omega)\otimes \C^4 \equiv
[L^2(\Omega)]^4 $. Under quite general conditions the spectrum of
$\mathbb{P}_0 $ coincides with $[0,\infty)$, and the perturbation $W$
introduces negative eigenvalues $\lambda_n(\pw(\Omega)).$ The Dirac operator
$\mathbb{D}_0 $ has typically a spectrum  $\R \setminus (-1,1) $ and $V$
causes eigenvalues $\lambda_n(\dv(\Omega))$ to appear in the gap
$(-1,1)$. Our concern in this paper is with the Riesz means
\begin{equation}\label{N3}
M_{\gamma}(\pw,\Omega) = \sum_n|\lambda_n(\pw(\Omega))|^{\gamma},\
\ \
M_{\gamma}(\dv,\Omega) = \sum_n|\lambda_n(\dv(\Omega))|^{\gamma}
\end{equation}
where $\gamma \ge 0 $, and, in particular, the counting functions
given by
\begin{equation}\label{N4}
N(\pw,\Omega)=M_0(\pw,\Omega),\ \ \ N(\dv,\Omega)=M_0(\dv,\Omega).
\end{equation}

In \cite{Sobolev1998}, Sobolev investigated the natural quasi-classical formula
\begin{equation}\label{N5}
M_{\gamma}(\pw,\Omega) \sim \hbar^{-3}\mathfrak{B}_
\gamma(\mu \hbar |\B|,W,\Omega), \hbar \rightarrow 0,
\end{equation}
with the ``magnetic" Weyl coefficient
\begin{equation}\label{N6}
\mathfrak{B}_\gamma(b,v,\Omega):= \beta_\gamma\int_\Omega
b(x)\left [v_-(x)^{\gamma +\frac12}+2\sum_{k=1}^\infty
[2kb(x)+v(x)]_-^{\gamma+\frac12}\right ]dx,
\end{equation}
where
\[
\beta_{\gamma} = \frac{1}{4\pi^2}\int_0^1t^{\gamma}(1-t)^{-1/2}dt.
\]
Under weak regularity conditions on $\B$ and $W$, Sobolev
established two main results:
\begin{enumerate}
\item [{($R_1$):}]
for $ \mu \hbar \le$ constant, (\ref{N5}) is satisfied for $\gamma
\ge 1$;
\item [{($R_2$):}]
if $\B$ has constant direction, (\ref{N5}) is satisfied uniformly in
$\mu \ge 0$ for $ \gamma >1/2$.
\end{enumerate}
Earlier results of this type were proved for homogeneous (i.e.
constant) fields by Lieb, Solovej, and Yngvason~\cite{Liebetal1994}, \cite{Liebetal1995},
and for non-homogeneous fields by Erd\H{o}s and Solovej~\cite{ErdosSolovej1999}, \cite{ErdosSolovej1997},
with $\mu \hbar^3 \rightarrow 0$. Our objective in this
paper is to investigate the validity of ($R_1$) and ($R_2$) for all
$\gamma \ge 0$. We prove in Theorem~\ref{Count} that if $W, |\B| \in
L^{3/2}(\Omega) $ in ($R_1$), then (\ref{N5}) is satisfied for all $\gamma
\ge 0$. This extends a result in \cite{EvansLewis1999} in which a Weyl asymptotic
formula is established for the case $\mu \hbar \sim 0 $ as $\hbar
\rightarrow 0$. If $|\mathbf{B}|$ does not belong to $ L^{3/2}(\Omega)$
in ($R_1$), we obtain the result for
$N(\pw+\lambda,\Omega)$, the number of eigenvalues of
$\pw(\Omega)$ less than $-\lambda <0$. Our main result in problem ($R_2$) is also
of this form. This is the best that can
be expected in general for $\gamma =0 $ in view of the absence of an
inequality of Cwikel, Lieb, Rozenblum type for the number of negative
eigenvalues; such an inequality is available when $| \mathbf{B}|
\in L^{3/2}(\Omega)$. We also investigate circumstances in which (\ref{N5}) holds
for $\gamma > 0$ when $ \mathbf{B}$ has a constant direction.

In \cite{Raikov1999} the leading term in the asymptotic value of
$N(\pw+\lambda,\R^3)$ is determined for $\lambda > 0, \hbar =1, \mathbf{B}$ of constant
direction, and $\mu
\rightarrow \infty$. Specifically, with $\B({\bf{x}})
= (0,0,B(x)), {\bf{x}}=(x,x_3), x\in \R^2,$ where $B(x)$ is bounded
above and away from zero and has a bounded gradient,
\begin{equation}\label{N7}
\lim_{\mu \rightarrow \infty}\mu^{-1}N(\pw+\lambda,\R^3)={\mathcal{D}}
(\lambda) :=
\frac{1}{2\pi}\int_{\R^2}N(X_{W+\lambda}(x),\R)B(x)dx,
\end{equation}
where, for any fixed $x\in \R^2, X_W(x)$ is a self-adjoint realization of
$-\frac{d^2}{dx_3^2}+W(x,\cdot)$ in $L^2(\R)$ with essential
spectrum $[0,\infty)$. From many other results on the spectral
asymptotics of $N(\pw,\Omega)$, we mention in particular that of
Iwatsuka and Tamura in \cite{IwatsukaTamura1998} for $\hbar=\mu=1$ and $\lambda
\rightarrow 0$ :
\begin{equation}\label{N8}
N(\pw+\lambda,\R^3) \sim \mathcal{D}(\lambda).
\end{equation}
Results for the two-dimensional problems are also obtained by the
authors cited above, and our techniques also apply in this case.
Furthermore, we derive analogous results for the Dirac
operator with magnetic field in both cases ($R_1$) and ($R_2$) above.

Our strategy is based on that in \cite{Sobolev1998} which in turn was inspired by
ideas from \cite{ColindeVerdiere1986}. For the case $\mu \hbar \le $constant, the
technique we use involves a tesselation of $\R^3$ by cubes $Q$,
the derivation of two-sided estimates for $N(\pw,Q)$, with constant
$W$ and $\B$, in terms of the explicit values of the eigenvalues
(Landau levels) of the operator realization of $\pw$ on a torus,
and subsequently the localization of $\pw(\Omega)$ in terms of the
$\pw(Q)$ and the application of either the Cwikel, Lieb, Rozenblum (CLR)
inequality for the magnetic Schr\"{o}dinger operator (in the case in which
there is known to be a finite number of negative eigenvalues in problem ($R_1$)),
or an inequality derived from a Lieb-Thirring inequality established by
Shen in \cite{Shen1999} for the trace of the negative eigenvalues.  For
problem ($R_2$) this method is too crude. Sobolev in \cite{Sobolev1998} used the
spectral properties of $\pw$ on a torus with an
arbitrary periodic magnetic field $\B$ having integer flux, and, in
particular, that zero is an eigenvalue of the operator with multiplicity
equal to the flux of $\B$. Another important ingredient in \cite{Sobolev1998} in this case
is a Lieb-Thirring inequality established in \cite{Sobolev1996} in which $|\mu
\B|$ has the proper (linear) scale; for a discussion of the significance of this see
the paper of Erd\H{o}s and Solovej~\cite{ErdosSolovej1999}, \S1. We use a similar
estimate derived by
Shen in \cite{Shen1999} for $N(\pw +\lambda,\R^3)$ with $\lambda >0$. This grows like
$1/\sqrt \lambda$ as $ \lambda \rightarrow 0 $, and implies (\ref{N5}) only when
$\gamma >1/2 $, as already established in \cite{Sobolev1998} with similar assumptions.
We also present a result for a cylindrical domain $\Omega$
in which $ N(\pw+\lambda, \Omega) $ grows
logarithmically, which implies that $M_{\gamma}(\pw,\Omega)$ is finite
for all $ \gamma >0$ (see Proposition~\ref{Prop5}).

Throughout the paper $(\cdot,\cdot)_{\Omega}$, $\|\cdot\|_{\Omega} $ are used to
denote the usual inner-product and norm in each of the Hilbert
spaces $L^2(\Omega)$, $[L^2(\Omega)]^2$ and $[L^2(\Omega)]^4 $; the precise space
will be clear from the context. We adopt the convention that
inner-products are linear in the second argument and conjugate
linear in the first. We denote the identity matrix on $ {\mathbb{C}}^4 =
{\mathbb{C}}^2 \bigotimes {\mathbb{C}}^2 $ by $ {\mathbb{I}}_2$. We denote
points in $\R^3$ as $\mathbf{x}=(x,x_3), x\in \R^2,x_3\in \R.$ We shall write
$A\lsim B$, or $B\gsim A $, to mean that $|A| \le CB$ for some positive constant
$C$, and $f(x) = O(A) $ will mean that $f(x)$ is bounded by $A$.

\section{Preliminaries}

To begin, we need to define the operators precisely and locate
their essential spectra. The Pauli operator $\pw$ will be defined
as the form sum of $\mathbb{P}_0$ and the operator of
multiplication by $W$, the conditions assumed on $W$ being
sufficient for the associated sesquilinear form to be lower
semi-bounded. For the Dirac operator we require the restriction of
$\dv$ to $C_0^{\infty}(\Omega)$ to be essentially self-adjoint, in
which case $\dv^2$ is the self-adjoint operator associated with
the form $\|\dv\phi\|_{\Omega}^2$. In \S3
below, our requirements in Theorem~\ref{Count} are met by the following results from
\cite{EvansLewis1999}. Define, for $m=2,4,$
$$
\mathcal{H}_a^1 \equiv \mathcal{H}_a^1(\R^3):=
\{u:u,((\hbar/i)\partial_{\nu}-\mu a_{\nu})u \in [L^2(\R^3)]^m,
\nu=1,2,3,\}
$$
in which the derivatives are defined in the distributional sense.
If $ \mathbf{a}\in [L^2_{loc}]^3$, then $[C_0^{\infty}(\R^3)]^m$ is dense
in $ \mathcal{H}_a^1$ (see Kato~\cite{Kato1978}, Simon~\cite{Simon1979},
and Leinfelder and Simader~\cite{LeinfelderSimader1981}).
Also, the diamagnetic inequality
\[
|(\hbar/i)\partial_{\nu}(|u|)|\le |((\hbar/i)\partial_{\nu}
-\mu a_{\nu})u|,\q \nu=1,2,3,
\]
holds for almost every $x\in\R^3$ and $u \in \mathcal{H}_a^1$ (see
Lieb and Loss~\cite{LiebLoss1997}, p.179). As a consequence, $u\mapsto|u|$ maps $\mathcal{H}_a^1
$ continuously into the Sobolev space $H^1(\R^3)$, which implies the existence of
a continuous embedding
\[
\mathcal{H}_a^1\hookrightarrow [L^s(\R^3)]^m,\q s\in[2,6]
\]
(see Edmunds and Evans~\cite{EdmundsEvans1987}, Theorem V.3.7). We denote by
$\mathcal{H}^1_{a,0}(\Omega)$ the closure of
$[C_0^{\infty}(\Omega)]^m$ in $ \mathcal{H}_a^1.$ The following
propositions follow in a similar way to Lemma 2.1 and Lemma 2.2 in \cite{EvansLewis1999} and
help to satisfy our needs in Theorem~\ref{Count}.

\begin{Proposition}\label{Prop1}
Let $a_{\nu} \in L^2_{loc}(\Omega)$, $\nu=1,2,3,$ and $W$,
$|\mathbf{B}|\in L^{3/2}(\Omega)$. Then $\pw = \mathbb{P}_0+W$ is
defined as a form sum with form domain
$\mathcal{H}^1_{a,0}(\Omega)$.
\end{Proposition}

\begin{Proposition}\label{Prop2}
Let $ \mathbf{a}$, $V (=W)$, $|\mathbf{B}|$ satisfy the conditions of
Proposition~\ref{Prop1} and suppose that the $a_{\nu}$ are locally Lipschitz
on $\Omega$. Then $\dv$ is essentially self-adjoint on
$[C_0^{\infty}(\Omega)]^4$. If furthermore $V\in L^3(\Omega)$, then
$\dv$ has domain $ \mathcal{H}^1_{a,0}(\Omega)$.
\end{Proposition}

At the end of \S2 of \cite{EvansLewis1999} conditions, consistent with those in
Proposition~\ref{Prop2}, are given which ensure that $\dv$ has essential spectrum
$ \mathbb{R}\setminus (-1,1) $.

The assumptions made in Propositions \ref{Prop1} and \ref{Prop2} are in fact
sufficient for there to exist only a finite number of negative
eigenvalues of $\pw(\Omega)$ (and in $(-1,1)$ for $\dv(\Omega)$).
To consider cases in which $| \mathbf{B}|$ is not in $ L^{3/2}(\Omega)$ we
make use of an estimate given by Shen in \cite{Shen1999} for the trace
$M_1(\pw,\mathbb{R}^3)$. The
assumptions that this requires are sufficient to ensure that $\pw$ is defined as a
form sum. To be specific, let
\[
l_p( \mathbf{x}):=\sup\left\{l>0:l^2\left(\frac{1}{l^3}
   \int_{Q(\mathbf{x},l)}|B(\mathbf{y})|^pd\mathbf{y}\right)^{1/p}\le 1\right\},
\]
where $Q( \mathbf{x},l)$ denotes the cube in $ \mathbb{R}^3$
center $\mathbf{x}$ and side $l$. Define
\begin{equation}\label{bp}
b_p(\mathbf{x}):=\frac{1}{[l_p( \mathbf{x})]^2}.
\end{equation}
This is Shen's ``effective" magnetic field. We recall that the need
to replace $| \mathbf{B}|$ by some screened version in
Lieb-Thirring inequalities for Pauli operators with non-homogeneous
magnetic fields was demonstrated by Erdos in \cite{Erdos1995}, and this had
motivated Sobolev in \cite{Sobolev1996} to obtain Lieb-Thirring estimates
similar in form to that obtained for constant fields in Lieb, Solovej, and
Yngvason~\cite{Liebetal1994} but with $ |\mathbf{B}|$ replaced by an ``effective"
magnetic field. Another result of this kind is established in
\cite{Bugliaroetal1997}. The novelty of Shen's approach lies in the simple and
natural way in which his $b_p$ is constructed. In \cite{Shen1999}, Remark
1.4, the following result is proved for $\Omega = \R^3$, but the
proof is valid in general. It also holds if $b_p$ is Sobolev's
effective magnetic field, or that of Bugliaro {\it et al}.

\begin{Proposition}\label{Prop3}
Let $ \mathbf{a}\in L^2_{loc}(\Omega, \mathbb{R}^3)$ and
suppose that for any $p>3/2$ and $\gamma \ge 1$, $W_-^{\gamma+3/2}$,
$b_p^{3/2}W_-^{\gamma} \in L^1(\Omega)$. Then $\pw$ is defined as a
form sum with form domain $ \mathcal{H}_{a,0}^1$.
\end{Proposition}

In the case when $ \mathbf{B}$ has constant direction, there is
another result of Shen~\cite{Shen1999} which fits our purpose, namely his estimate for the
counting function $N(\pw +\lambda, \mathbb{R}^3)$. Let
$ \mathbf{B}(\mathbf{x})=(0,0,B(x))$, where $ \mathbf{x}=(x,x_3),
x\in \mathbb{R}^2$, and define
\[
l_p(x):=\sup\left\{l>0:l^2\left(\frac{1}{l^2}
   \int_{S(x,l)}|B(y)|^pdy\right)^{1/p}\le 1\right\},
\]
where $S(x,l)$ denotes the square in $ \mathbb{R}^2 $ center
$x$ and side $l$. Define
\begin{equation}\label{bphat}
\hat{b}_p(x):= \frac{1}{[l_p(x)]^2}.
\end{equation}
The following result is proved in [\cite{Shen1999}, Remark 1.4].

\begin{Proposition}\label{Prop4}
Let $ \mathbf{a}\in L^2_{loc}(\Omega, \mathbb{R}^3)$ and
suppose that for any $p>1$ and $\gamma > 1/2$, $W_-^{\gamma+3/2}$,
$\hat{b}_pW_-^{\gamma+1/2} \in L^1(\Omega)$. Then, with
$\mathbf{B}=(0,0,B)$, $\pw$ is defined as a
form sum with form domain $ \mathcal{H}_{a,0}^1$.
\end{Proposition}

In the rest of this section we present preparatory results for
subsequent sections. Our initial assumptions are as follows:
\begin{itemize}
\item[($A_1$)] Let $Q =\cup_{k=1}^KQ_k$, where
$\{Q_k\subseteq \Omega :k=1,\dots,K\}$ is a finite collection
of non-overlapping congruent cubes whose edges are parallel to the
coordinate axes and of length $r$. The set $Q$ is fixed, but the side
lengths $r$ and number $K$ will depend on $\hbar$ in due course.
\item[($A_2$)] Let ${V_0}$ and $W_0$ be piecewise constant functions, taking
constant values in each $Q_k$, and zero outside $Q$.
\item[($A_3$)] Assume that $\B$ is continuous. Define $\B^o$ on $Q$ by $\B^o(\mathbf{x})
=\B(\mathbf{x}_k)$, $\mathbf{x}\in
Q_k$, where $\mathbf{x}_k$ is the center of the cube $Q_k$, and let $\B^o(\mathbf{x})\equiv
0$ for $\mathbf{x}\notin Q$. Choose\footnote{See the Appendix for details.} gauges
$\sa$ and $\AAAA$ for $\B$ and $\B^o$, respectively, such that for every
$Q_k$
\begin{equation}\label{Eqn2.4,5}
 \max_{\mathbf{x}\in Q_k}|\sa(\mathbf{x})-\AAAA(\mathbf{x})|\le
Cr\sigma_r,\q \sigma_r:=\max_{|\mathbf{x}-\mathbf{y}|<r}|\B(\mathbf{x})-\B(\mathbf{y})|.
\end{equation}
\end{itemize}

Define
\begin{equation}\label{Eqn2.3}
\mathfrak{B}_\gamma(b,v,\Omega):= \beta_\gamma\int_\Omega
b(\mathbf{x})\left [ v_-(\mathbf{x})^{\gamma +\frac12}+2\sum_{k=1}^\infty
[2kb(\mathbf{x})+v(\mathbf{x})]_-^{\gamma+\frac12}\right ] d\mathbf{x}
\end{equation}
for $\gamma \ge 0$ where
$$
\beta_\gamma:=\frac{1}{4\pi^2}\int_0^1(1-t)^{-\frac12} t^{\gamma}dt. $$
For simplicity, we shall write $\mathfrak{B}(b,v,\Omega)$ for
$\mathfrak{B}_0(b,v,\Omega)$ when there is no danger of confusion. We are
interested in the asymptotic behavior of the Riesz means
$$
\sum_k|\lambda_k|^\gamma,\ \ \ \gamma \ge 0,
$$
where each $\lambda_k$ is a negative eigenvalue of the Pauli operator
$\mathbb{P}_W(\Omega)$, or each $\lambda_k$ is an eigenvalue of the Dirac operator $\dv$ in
$(-1,1)$.

We make frequent use of Proposition~3.2 of Sobolev~\cite{Sobolev1998},
which follows from results of Colin de
Verdiere~\cite{ColindeVerdiere1986}.
It states that there exists a constant $C$ such that for any
$\delta\in (0,\frac12)$
\begin{equation}\label{Eqn2.1}
\hbar^3N(\mathbb{P}_0(\B^o) +\lambda,Q_k)\le
\mathfrak{B}(\mu\hbar|\B^o|,\lambda,Q_k)
\end{equation}
and
\begin{equation}\label{Eqn2.2}
\hbar^3N(\mathbb{P}_0(\B^o)+\lambda,Q_k)\ge
(1-\delta)^3
\mathfrak{B}(\mu\hbar|\B^o|,\lambda+\frac{C\hbar^2}{\delta^2r^2},Q_k).
\end{equation}
Here $\mathbb{P}_0(\B^o)(Q_k)$ is the Dirichlet operator on $Q_k$,
and $ N( {\mathbb{P}}_0( \mathbf{B}_0)+\lambda,Q_k)$ the number of its
negative eigenvalues below $ -\lambda <0$.

\begin{Lemma}\label{Lem2.1} Define
$\sa^\infty:= \sa-\AAAA.$
For all $\phi\in [\Con(Q_k)]^4$ and $\theta >0$
\begin{equation}\label{Eqn2.4}\begin{array}{l}
|([(\dvo(\B)-{V_0})^2-1+W_0
-(\mathbb{P}_0(\B^o)+W_0)\mathbb{I}_2]\phi,\phi)_{Q_k}|\\
\le
\theta(\mathbb{P}_0(\B^o)\mathbb{I}_2\phi,\phi)_{Q_k}+\mu^2(1+\frac{1}
{\theta})
(|\sa^\infty|^2\phi,\phi)_{Q_k}.
\end{array}
\end{equation}
\end{Lemma}
\begin{proof} Set $D_\sa:=\frac{\hbar}{i}\grad -\mu\sa$ and
$D_0=D_{\AAAA}$.
We have that
$$\begin{array}{l}
([(\dvo-{V_0})^2-1+W_0]\phi,\phi)_{Q_k}\\
=([(\al\cdot D_\sa +\beta)^2-1+W_0]\phi,\phi)_{Q_k}\\

=\|[\al\cdot (D_0 -\mu\sa^\infty)\phi\|^2_{Q_k}+(W_0\phi,\phi)_{Q_k}\\

= \|\al\cdot D_0 \phi\|_{Q_k}^2 -
2\mu\mathfrak{Re}[([\al\cdot D_0]\phi,[\al\cdot\sa^\infty]\phi)_{Q_k}]
+\mu^2\|[\al\cdot\sa^\infty]\phi\|^2_{Q_k}+(W_0\phi,\phi)_{Q_k}\\

= (\mathbb{P}_{W_0}(\B^o)\mathbb{I}_2\phi,\phi)_{Q_k} -
2\mu\mathfrak{Re}[([\al\cdot D_0]\phi,[\al\cdot\sa^\infty]\phi)_{Q_k}]
+\mu^2\|[\al\cdot\sa^\infty]\phi\|^2_{Q_k}
\end{array}
$$
since $\|\al\cdot
D_0 \phi\|_{Q_k}^2=(\mathbb{P}_0(\B^o)\mathbb{I}_2\phi,\phi)_{Q_k}$.
Therefore,
$$\begin{array}{l}
\big |\big
([(\dvo-{V_0})^2-1+{W_0}-\mathbb{P}_{W_0}(\B^o)\mathbb{I}_2]\phi,\phi\big
)_{Q_k}\big |\\
\le \theta (\mathbb{P}_0(\B^o)\mathbb{I}_2\phi,\phi)
+(1+\frac{1}{\theta})\mu^2 \|[\al\cdot\sa^\infty]\phi\|_{Q_k}^2
\end{array}
$$ which completes the proof.
\end{proof}

It follows from Lemma~\ref{Lem2.1} and (\ref{Eqn2.4,5}) that for
$\phi\in[\Con(Q_k)]^4$
and $\theta\in (0,1)$
\begin{equation}\label{Eqn2.5}\begin{array}{l}
([(\dvo(\B)-{V_0})^2-1+{W_0}]\phi,\phi)_{Q_k}\le \\
\ \ \  (1+\theta)(\mathbb{P}_0(\B^o)\mathbb{I}_2\phi,\phi)_{Q_k} +
{W_0}\|\phi\|_{Q_k}^2
+\frac{C^2}{\theta}\mu^2r^2\sigma_r^2\|\phi\|^2_{Q_k}\end{array}
\end{equation}
and
\begin{equation}\label{Eqn2.6}\begin{array}{l}
([(\dvo(\B)-{V_0})^2-1+{W_0}]\phi,\phi)_{Q_k}\ge \\
\ \ \  (1-\theta)(\mathbb{P}_0(\B^o)\mathbb{I}_2\phi,\phi)_{Q_k} +
{W_0}\|\phi\|_{Q_k}^2
-\frac{C^2}{\theta}\mu^2r^2\sigma_r^2\|\phi\|^2_{Q_k}.\end{array}
\end{equation}
Hence, from (\ref{Eqn2.1}) and (\ref{Eqn2.2}) we have that
\begin{equation}\label{Eqn2.7}\begin{array}{l}
\frac12\hbar^3 N([(\dvo(\B)-{V_0})^2-1+{W_0}];Q_k)\ge \\
\ \ \  (1-\delta)^3
\mathfrak{B}(\mu\hbar|\B^o|,\frac{{W_0}}{1+\theta}+\frac{C^2}{\theta(1+
\theta)}\mu^2r^2\sigma_r^2
+\frac{C\hbar^2}{\delta^2r^2};Q_k)
\end{array}
\end{equation}
and
\begin{equation}\label{Eqn2.8}
\begin{array}{l}
\frac12\hbar^3 N([(\dvo(\B)-{V_0})^2-1+{W_0}];Q_k)\le \\
\ \ \ \mathfrak{B}(\mu\hbar|\B^o|,
\frac{{W_0}}{1-\theta}-\frac{C^2}{\theta(1-\theta)}\mu^2r^2\sigma_r^2,Q_k).
\end{array}
\end{equation}
Similar estimates follow for the Pauli operator:
\begin{equation}\label{N20}
\begin{array}{l}
\hbar^3 N(\mathbb{P}_{W_0}(\B);Q_k)\ge \\
\ \ \  (1-\delta)^3
\mathfrak{B}(\mu\hbar|\B^o|,\frac{W_0}{1+\theta}+\frac{C^2}
{\theta(1+\theta)}\mu^2r^2\sigma_r^2
+\frac{C^2\hbar^2}{\delta^2r^2},Q_k)
\end{array}
\end{equation}
and
\begin{equation}\label{Eqn2.9B}
\begin{array}{l}
\hbar^3 N(\mathbb{P}_{W_0}(\B);Q_k)\le \\
\ \ \ \mathfrak{B}(\mu\hbar|\B^o|,
\frac{W_0}{1-\theta}
-\frac{C^2}{\theta(1-\theta)}\mu^2r^2\sigma_r^2,Q_k)
\end{array}.
\end{equation}
To estimate error terms, we use the following inequalities for
$ \mathfrak{B} \equiv \mathfrak{B}_0 $ established in
Sobolev~\cite{Sobolev1998}: for any $\lambda \ge 0$ and any subset
$G$ of $ \Omega $
\begin{equation}\label{Eqn2.10}
\begin{array}{l}
|\mathfrak{B}(b,U_1+ \lambda,G)-\mathfrak{B}(b,U_2 +\lambda,G)| \\
\ \ \ \lsim \mathcal{M}(b,U_1-U_2,G) +
\int_G|U_1-U_2|^{1/2}(|U_1|+|U_2|) d \mathbf{x}\\
\ \ \ \lsim \mathcal{M}(b,U_1-U_2,G) + \mathcal{N}(U_1-U_2,G)^\frac13
\{\mathcal{N}(U_1-U_2,G)^\frac23
+ \mathcal{N}(U_2,G)^\frac23\},\end{array}
\end{equation}
and
\begin{equation}\label{Eqn2.11}\begin{array}{l}
|\mathfrak{B}(b_1,U,G)-\mathfrak{B}(b_2,U,G)| \lsim \\
\ \ \ \mathcal{M}(|b_1-b_2|,U,G)
+\mathcal{M}(|b_1-b_2|,U,G)^\frac12 \mathcal{N}(U,G)^\frac12\\
\ \ \ \ + \mathcal{M}(|b_1-b_2|,U,G)^\frac14 \mathcal{N}(U,G)^\frac34 +
\mathcal{M}(|b_1-b_2|,U,G)^\frac12 \mathcal{N}(U,G)^\frac12
\end{array}
\end{equation}
where
$$
\mathcal{M}(b,U,G):= \int_G b(x)|U(x)|^\frac12 dx,\ b\ge 0,\q \mathcal{N}(U,G):=
\int_G|U(x)|^\frac32 dx.
$$
\begin{Lemma}\label{Lem2.2} If  $r = A\hbar$, then
\begin{equation}\label{Eqn2.12}
\begin{array}{l} |Q_k|^{-1}\left
|\mathfrak{B}(\mu\hbar|\B^o|,\frac{W_0}{1+\theta} +
C_1\frac{\mu^2r^2\sigma_r^2}{\theta(1+\theta)} +
C_2\frac{\hbar^2}{\delta^2r^2},Q_k) -
\mathfrak{B}(\mu\hbar|\B|,W_0,Q_k)\right|\\ \lsim
\mu\hbar|\B^o|(\theta
+\frac{A^2}{\theta}\mu^2\hbar^2\sigma_r^2+\frac{1}{A^2\delta^2})^\frac12 +
(\theta
+\frac{A^2}{\theta}\mu^2\hbar^2\sigma_r^2+\frac{1}{A^2\delta^2})^\frac32 +
(\mu\hbar\sigma_r)^\frac14 + \mu\hbar\sigma_r.
\end{array}
\end{equation}
\end{Lemma}
\begin{proof} With $U_2:=W_0$ and
$$
U_1:=\frac{W_0}{1+\theta} +
C_1\frac{\mu^2r^2\sigma_r^2}{\theta(1+\theta)} +
C_2\frac{\hbar^2}{\delta^2r^2}, $$
$$\begin{array}{rl} |U_1-U_2| \le &
|W_0|\theta +C_1\frac{\mu^2r^2\sigma_r^2}{\theta(1+\theta)} +
C_2\frac{\hbar^2}{\delta^2r^2}\\ \lsim & \theta
+\frac{A^2}{\theta}\mu^2\hbar^2\sigma_r^2 +
\frac{1}{A^2\delta^2}=:F(\theta, A,\mu\hbar,\delta,\sigma_r)
\end{array}
$$
and
$$ |\B-\B^o|\le \sigma_r,\q x\in Q_k.
$$
Therefore,
$$
\mathcal{M}(\mu\hbar|\B-\B^o|,{W_0},Q_k)\lsim \mu\hbar\sigma_r|Q_k|, $$ $$
\mathcal{M}(\mu\hbar|\B^o|,U_1-U_2,Q_k)\lsim
\mu\hbar|\B^o|F(\theta,A,\mu,\hbar,\delta,\sigma_r)^\frac12 |Q_k|,\q
\mathcal{N}(U_2,Q_k)\lsim |Q_k|,
$$
and
$$ \mathcal{N}(|U_1-U_2|,Q_k)\lsim
F(\theta,A,\mu,\hbar,\delta,\sigma_r)^\frac32 |Q_k|.
$$
The lemma follows from (\ref{N20}) and (\ref{Eqn2.9B}).
\end{proof}

\section{The Pauli operator: $\mu\hbar\le constant$.}
We assume that ($A_1$)-($A_3$) hold in \S2 with ${V_0}\equiv 0$. It follows
that $\mathbb{P}_{W_0}(\Omega)\le \oplus_{k=1}^K\mathbb{P}_{W_0}(Q_k)$
in the form sense (see [\cite{EdmundsEvans1987},\S XI.2.2]), which implies that
\begin{equation}\label{Eqn3.1}
N(\mathbb{P}_{W_0}, \Omega)\ge \sum_{k=1}^K
N(\mathbb{P}_{W_0}, Q_k),
\end{equation}
and, since
\begin{equation}\label{N26}\begin{array}{rl}
M_{\gamma}( \mathbb{P}_{W_0}+\lambda,\Omega) = &
-\int_0^{\infty} t^{\gamma}dN(\mathbb{P}_{W_0}+\lambda+t,\Omega)\\
\ \ & \ \ \\
= & \gamma
\int_0^{\infty} t^{\gamma-1}N(\mathbb{P}_{W_0}+\lambda+t,\Omega)dt,
\end{array}
\end{equation}
for all $\lambda>0$,
we have for all $\gamma \ge 0$
\begin{equation}\label{N27}
M_{\gamma}(\mathbb{P}_{W_0}+\lambda,\Omega)\ge \sum_{k=1}^K
M_{\gamma}(\mathbb{P}_{W_0}+\lambda,Q_k).
\end{equation}
Note that (\ref{N26}), and hence (\ref{N27}), hold for $\lambda
=0$ if $N(\mathbb{P}_{W_0},\Omega)<\infty$, which will be the case
in Theorem~\ref{Count} below, but not in Theorem~\ref{Count2} and
Theorem~\ref{Thm5.6}. In Theorem~\ref{Count2} and
Theorem~\ref{Thm5.6} all we know is that
$N(\mathbb{P}_{W}+\lambda,\Omega)=O(\lambda^{-1})$ and
$O(\lambda^{-\frac12})$ respectively, and as a consequence, we can
claim that (\ref{N26}) holds for $\lambda=0$ only if $\gamma >1$
in the first case and $\gamma >\frac12$ in the second.
We are mainly interested in the remaining values of $\gamma$ in
Theorems \ref{Count2} and \ref{Thm5.6}, and take $\lambda >0$.

Also, note that $\mathbb{P}_{W_0} \ge -(W_0)_- \ge -\Lambda $, say, and the
inequalities (\ref{N20}) and (\ref{Eqn2.9B}) with $W_0$ replaced by
$W_0+\lambda$ hold uniformly for $\lambda \in [0,\Lambda]$. Moreover,
\begin{equation}\label{Bgamma}
\mathfrak{B}_{\gamma}(b,W,G) =
\gamma \int_0^{\infty}{t}^{\gamma-1}\mathfrak{B}_0(b,W+t,G)dt.
\end{equation}
Hence, from (\ref{Eqn2.7}), (\ref{Eqn2.12}), and
(\ref{N26}), we see that when $\mu\hbar\lsim 1$
$$\begin{array}{l}
\hbar^3 M_{\gamma}(\mathbb{P}_{W_0}+\lambda;\Omega)\ge
(1-\delta)^3\mathfrak{B}_{\gamma}(\mu\hbar|\B|,W_0+\lambda,Q)\\
\ \ \ -O\left (\left\{(\theta
+\frac{A^2}{\theta}\sigma_r^2+\frac{1}{A^2\delta^2})^\frac12 +
(\theta +\frac{A^2}{\theta}\sigma_r^2
+\frac{1}{A^2\delta^2})^\frac32 + \sigma_r^\frac14
+\sigma_r\right\}|Q|\right ).
\end{array}
$$
On allowing $\hbar\to 0$, $\theta\to 0$, $A\to\infty$, and
$\delta\to 0$ in that order we have that for all $\gamma \ge0$
\begin{equation}\label{Eqn3.2}
\liminf_{\hbar\to 0}\{\hbar^3M_{\gamma}(\mathbb{P}_{W_0}+\lambda;\Omega)
-\mathfrak{B}_{\gamma}(\mu\hbar|\B|,W_0+\lambda,\Omega)\}\ge 0.
\end{equation}
To prove the reverse inequality in (\ref{Eqn3.2}), we proceed in
the manner of Y. Colin de Verdiere~\cite{ColindeVerdiere1986} and
Sobolev~\cite{Sobolev1998}.
Let the interior of each $Q_k$ be denoted by
$int (Q_k)$. Set
\begin{equation}\label{Eqn3.3}
S:=\R^3 \setminus \cup_{k=1}^K int (Q_k),\q  {S}_{\rho r} :=\{x\in\R^3:
dist(x,S)<\rho r\}
\end{equation}
for some $\rho\in (0,1)$.
 Construct a partition of unity  $\{\psi_k\}_{k=0}^K$ subordinate
 to the covering $\cup_{k=1}^K int (Q_k)\cup S_{\rho r}$ of $\R^3$:
\begin{equation}\label{Eqn3.4}\begin{array}{rl}
(i) & \psi_0\in C^\infty(\R^3),\ \ \psi_k\in \Con(int (Q_k)), k\ge 1,\\
(ii) & \sum_{k=0}^K\psi_k^2 \equiv 1,\\
(iii) & \sum_{k=0}^K |\grad \psi_k(x)|^2
\lsim (\rho r)^{-2}\chi_{\rho r},\end{array}
\end{equation}
where $\chi_{\rho r}$ is the characteristic function of
${Q}_{\rho r}:=\{x\in\R^3:dist(x,{Q})<\rho r\}$. Note that
\begin{equation}\label{Eqn3.5}
|S_{\rho r}\cap Q|\lsim \rho |Q|,\q |Q_{\rho r}\setminus Q|\lsim \rho r.
\end{equation}
Then, for every $f\in [\Con(\Omega)]^2$, a calculation yields
$$
(\mathbb{P}_0\psi_0f,\psi_0f)+\sum_{k=1}^K(\mathbb{P}_0\psi_kf,\psi_kf)
\le (\mathbb{P}_0 f,f) +C\hbar^2(\rho r)^{-2}(\chi_{\rho r}f,f)
$$
and, as a consequence,
\begin{equation}\label{Eqn3.6}\begin{array}{rl}
(\mathbb{P}_{W_0}(\Omega)f,f)\ge & ([\mathbb{P}_{W_0}({S}_{\rho
r}\cap\Omega)-C\hbar^2(\rho r)^{-2}\chi_{\rho r}]\psi_0f,\psi_0f)\\
&+\sum_{k=1}^K([\mathbb{P}_{W_0}(Q_k)-C\hbar^2(\rho r)^{-2}\chi_{\rho
r}]\psi_k f,\psi_k f).
\end{array} \end{equation}
By the minimax principle, $\lambda_n(\psi_k\mathbb{P}_{W_0}(Q)\psi_k)\ge
\lambda_n(\mathbb{P}_{W_0}(Q))$ for $k=0,1,\dots,K,$ (see (21) of
\cite{Evansetal1996b}). Therefore, it follows from (\ref{Eqn2.12}) and (\ref{Eqn3.6}) that
for all $\gamma \ge 0$
\begin{equation}\label{Eqn3.7}\begin{array}{rl}
M_{\gamma}(\mathbb{P}_{W_0}+\lambda,\Omega)\le &
M_{\gamma}(\mathbb{P}_{W_0}+\lambda-C\hbar^2(\rho r)^{-2}\chi_{\rho r},{S}_{\rho
r}\cap\Omega)\\ & + \sum_{k=1}^K
M_{\gamma}(\mathbb{P}_{W_0}+\lambda-C\hbar^2(\rho r)^{-2}\chi_{\rho
r},Q_k).\end{array}
\end{equation}
From (\ref{Eqn2.9B}), we have in the case that $\mu\hbar\lsim 1$,
and uniformly for $t \in [0,\Lambda]$,
$$\begin{array}{l}
\hbar^3 N(\mathbb{P}_{W_0}+\lambda+t-C\hbar^2(\rho r)^{-2},Q_k) \\
\le \mathfrak{B}(\mu\hbar|\B^o|,
\frac{W_0+\lambda+t}{1-\theta}-\frac{C^2}{\theta(1-\theta)}\mu^2r^2\sigma_
r^2
-\frac{C\hbar^2}{(1-\theta)\rho^2 r^2},Q_k)\\
\le \mathfrak{B}(\mu\hbar|\B|,W_0+\lambda+t,Q_k) + O\Big (\big\{(\theta
+\frac{A^2}{\theta}\sigma_r^2 + \frac{1}{A^2\rho^2})^\frac12\\ + (\theta
+\frac{A^2}{\theta}\sigma_r^2 + \frac{1}{A^2\rho^2})^\frac32 +
(\sigma_r)^\frac14 + \sigma_r\big\}|Q_k|\Big )
\end{array}
$$
as in Lemma~\ref{Lem2.2}. Hence, on using (\ref{N26}),
\begin{equation}\label{Eqn3.8} \begin{array}{l}
\hbar^3 \sum_{k=1}^K M_{\gamma}(\mathbb{P}_{W_0}+\lambda-C\hbar^2(\rho
r)^{-2},Q_k) \\ \le \mathfrak{B}_{\gamma}(\mu\hbar|\B|,W_0+\lambda,Q) + O\Big
((\theta +\frac{A^2}{\theta}\sigma_r^2 +
\frac{1}{A^2\rho^2})^\frac12\\ + (\theta
+\frac{A^2}{\theta}\sigma_r^2 +
\frac{1}{A^2\rho^2})^\frac32 + \sigma_r^\frac14 + \sigma_r\Big
).\end{array} \end{equation}
For the first term on the right side of the inequality
(\ref{Eqn3.7}), we use the Cwikel-Lieb-Rozenblum inequality for a
Schr\"odinger operator with a magnetic field\footnote{See Theorem~2.15 of
Avron {\it et al.}~\cite{Avronetal1978} or Lemma~4.1 of
\cite{EvansLewis1999}.} to derive the estimate
\begin{equation} \label{Eqn3.9}
\begin{array}{l} \hbar^3
M_{\gamma}(\mathbb{P}_{W_0}+\lambda-C\hbar^2(\rho r)^{-2}\chi_{\rho r},{S}_{\rho
 r}\cap\Omega)\\
\ \ \lsim \int_{S_{\rho r}\cap\Omega}[\mu\hbar|\B|
-W_0+\frac{\hbar^2}{(\rho r)^2}\chi_{\rho r}]_{+}^{\frac32+\gamma}d\mathbf{x}\\ \
\  \lsim \int_{S_{\rho r}\cap \Omega}([|{W_0}| +|\B|]^{\frac32
+\gamma}
 +[\frac{\hbar}{\rho r}]^{2\gamma +3}\chi_{\rho r})d\mathbf{x}\\
\ \ \lsim
 \int_{\Omega\setminus Q}|\B|^{\frac32+\gamma}d\mathbf{x}
 + \int_{S_{\rho r}\cap Q}[|{W_0}|+|\B|]^{\frac32 +\gamma}d\mathbf{x}
 + [\frac{\hbar}{(\rho r)}]^{2\gamma +3}|Q_{\rho r}\cap S_{\rho r}|\\
\ \ \lsim \int_{\Omega\setminus Q}|\B|^{\frac32 +\gamma}d\mathbf{x}
 + K(\rho r) + \frac{1}{(A \rho)^{2\gamma +3}}[\rho _+\rho r]
 \end{array}
\end{equation}
where
$$
K(\rho r) = \int_{S_{\rho r}\cap Q}[|{W_0}| +
|\B|]^{\frac32 +\gamma}d\mathbf{x} \to 0,\q \text{as\ \ \ } \rho\to 0,
$$
uniformly in $\hbar$ and $A$, in view of (\ref{Eqn3.5}). From
(\ref{Eqn3.7}), (\ref{Eqn3.8}), and (\ref{Eqn3.9}), on allowing
$\hbar\to 0$, $\theta\to 0$, $A\to\infty$, and $\rho\to 0$, in that order,
it follows that
\begin{equation}\label{Eqn3.10}
\limsup_{\hbar\to 0}\left
\{\hbar^3 M_{\gamma}(\mathbb{P}_{W_0}+\lambda,\Omega)
-\mathfrak{B}_{\gamma}(\mu\hbar|\B|,{W_0+\lambda},\Omega)\right\} \lsim
\int_{\Omega\setminus Q}|\B(x)|^{\frac32 +\gamma} dx.
\end{equation}
Since $W_0=0$ outside $Q$, then $Q$ can be chosen such that
the right-hand side of (\ref{Eqn3.10})
is arbitrarily small if $| \mathbf{B}|\in L^{3/2+\gamma}(\Omega)$.
From this fact and (\ref{N27}), we conclude that for all $\gamma \ge 0$
\begin{equation}\label{Eqn3.11}
\lim_{\hbar\to 0}\left \{\hbar^3
M_{\gamma}(\mathbb{P}_{W_0}+\lambda;\Omega)
-\mathfrak{B}_{\gamma}(\mu\hbar|\B|,W_0+\lambda,\Omega)\right\} =0
\end{equation}
Now, we are in a position to prove the following general result for a Pauli
operator with an electric potential that is not assumed to be
piecewise constant.
\begin{Theorem}\label{Count}
Suppose that
\begin{itemize}
\item [(i)] $\B$ is continuous on $\Omega$,
\item [(ii)]  $W,|\B|\in L^{\frac32 +\gamma}(\Omega)$,
\item [(iii)] $\mu\hbar\ \le$ constant.
\end{itemize}
Then, for all $\gamma \ge 0$,
$$
\lim_{\hbar\to 0} \{\hbar^3 M_{\gamma}(\mathbb{P}_W,\Omega) -
\mathfrak{B}_{\gamma}(\mu\hbar |\B|,W,\Omega)\} = 0 $$
\end{Theorem}
\begin{proof} Given $\epsilon>0$, there exists a
finite collection of non-overlapping congruent cubes $\{Q_k\}_{k=1}^K$ for
which ($A_1$)-($A_3$) of \S2 hold (for $V_0\equiv 0$) with
\begin{equation}\label{Eqn3.12}
\|W-W_0\|_{\frac32+\gamma,\Omega}<\epsilon,\q
\|\B-\B^o\|_{\frac32+\gamma,\Omega}<\epsilon.
\end{equation}
Note that $W_0$ depends on $\varepsilon $ and hence so does the lower bound
$\Lambda $ of $\mathbb{P}_{W_0}$.
Let $\eta\in (0,1)$ and set
\begin{equation}\label{Eqn3.13}\begin{array}{rl}
\mathbb{P}_{W} =& T_1+T_2,\\
T_1:= & (1-\eta)\mathbb{P}_0 + W_0 -\eta\mu\hbar\bsig\cdot\B^o\\
T_2:= & \eta\mathbb{P}_0  + W_\infty +\eta\mu\hbar\bsig\cdot\B^o
\end{array}
\end{equation}
for $W_\infty:=W-W_0$. Then,
$$\begin{array}{rl}
T_1=&[\bsig\cdot(-i\tilde h\grad-\tilde\mu\sa)]^2
+W_0-\eta\mu\hbar\bsig\cdot\B^o\\ T_2=&\eta H_0(\B)+ W_\infty
-\eta\mu\hbar\bsig\cdot\B^\infty
\end{array}
$$
where
$$
\tilde h :=
(1-\eta)^\frac12\hbar,\ \ \tilde\mu:= (1-\eta)^\frac12\mu,\ \ \text{and\ \
} \B^\infty:=\B-\B^o.
$$
We have that
$$ T_1\ge S_1:= [\bsig\cdot(-i\tilde
h\grad -\tilde\mu\sa)]^2 + W_0 -\eta\mu\hbar|\B^o|
$$
and
$$
T_2\ge S_2:= \eta H_0(\B)+ W_\infty-\eta\mu\hbar|\B^\infty|.
$$
It then
follows  that
\begin{equation}\label{Weyl}
M_{\gamma}(\mathbb{P}_{W},\Omega)\le M_{\gamma}(S_1,\Omega)+
M_{\gamma}(S_2,\Omega).
\end{equation}
From (\ref{Eqn3.11}) with $\lambda =0$ (which, according to the
remark after (\ref{N27}), is allowed since $N(S_1,\Omega)<\infty$)
\begin{equation}\label{Eqn3.14}
\lim_{\hbar\to 0}\{\tilde
h^3 M_{\gamma}(S_1,\Omega)-\mathfrak{B}_{\gamma}(\tilde\mu\tilde h|\B|,W_0
-\eta\mu\hbar|\B^o|,\Omega)\}= 0.
\end{equation}
and by Theorem~2.15 of
Avron {\it et al.}~\cite{Avronetal1978}
\begin{equation}\label{Eqn3.15}
\hbar^3 M_{\gamma}(S_2,\Omega)\lsim \int_\Omega
[\eta^{-1}W_\infty-\mu\hbar|\B^\infty|]_-^{\frac32 +\gamma}dx\lsim
(\epsilon/\eta)^{\frac32 +\gamma}.
\end{equation}
Let $\gamma =0$. On using (\ref{Eqn2.10}), we have that
\begin{equation}\label{Eqn3.16}
\begin{array}{l}
|\mathfrak{B}(\tilde\mu\tilde h|\B|,W_0-\eta\mu\hbar|\B^o|,\Omega)
-\mathfrak{B}(\mu\hbar|\B|,W,\Omega)|\\
\lsim \int_\Omega |\B|[|W_\infty|+\eta|\B^o|]^\frac12 dx\\
\ \ \ + \|W_\infty +\eta|\B^o|\|_{\frac32,\Omega}^\frac12\{\|W_\infty
+\eta |\B^o|\|_{\frac32,\Omega} + \|W\|_{\frac32,\Omega}\}.
\end{array} \end{equation}
It follows from (\ref{Eqn3.13})-(\ref{Eqn3.16}) that
$$
\limsup_{\hbar\to 0}\{\hbar^3 N(\mathbb{P}_W;\Omega) -
\mathfrak{B}(\mu\hbar|\B|,W,\Omega)\}\le 0.
$$
The cases $\gamma >0$ follow similarly from (\ref{N26}) and the inequalities (2.23)
- (2.25) for $\mathbf{B}_{\gamma}$ in \cite{Sobolev1998}. The reverse
inequality is obtained by choosing $\eta \in(-1,0)$ and repeating
the argument.
\end{proof}

If $| \mathbf{B}|$ is not in $L^{3/2}(\Omega)$, there may be an infinite
number of negative eigenvalues. In this case we have the following
\begin{Theorem}\label{Count2}
Suppose that
\begin{enumerate}
\item
$ \mathbf{B} $ is continuous,
\item
$| \mathbf{B}|$, $W \in L^{\infty}(\Omega)$,
\item
for $p>3/2$, $W^{5/2}$, $b_p^{3/2}W \in L^1(\Omega) $, where $b_p$
is defined in (\ref{bp}),
\item
$\mu \hbar \le$ constant.
\end{enumerate}
Then, for all $\gamma \in [0, 1)$ and $\lambda >0$,
\begin{equation}\label{N45}
\lim_{\hbar \rightarrow 0}\left \{\hbar^3 M_{\gamma}(\pw +\lambda;
\Omega)- \mathfrak{B}_{\gamma}(\mu \hbar| \mathbf{B}|,W +\lambda,
\Omega)\right\} =0.
\end{equation}
If $\gamma \ge 1$, (\ref{N45}) with $\lambda =0$ is proved in \cite{Sobolev1998}.
\end{Theorem}
\begin{proof}
We first note that $\pw$ is properly defined as a form sum by
Proposition~\ref{Prop3}. Also, from [\cite{Shen1999}, Theorem 1.1] for $\lambda >0$,
\begin{eqnarray}\label{N46}
N(\pw +\lambda, \mathbb{R}^3)
&\le &|\lambda|^{-1}M_1(\pw,\mathbb{R}^3)\nonumber\\
& \lsim & |\lambda|^{-1}\{\hbar^{-3}\int_{\mathbb{R}^3}W_-^{5/2}d \mathbf{x}+
\mu^{3/2}\hbar^{-3/2}\int_{\mathbb{R}^3}b_p^{3/2}W_-d \mathbf{x}\}
\end{eqnarray}
and hence, for any $\gamma \ge0$ and $\lambda >0$
\begin{equation}\label{N47}
M_{\gamma}(\pw +\lambda, \Omega)
 \lsim |\lambda|^{-1}\{\hbar^{-3}\int_{\Omega}W_-^{5/2}d \mathbf{x}+
\mu^{3/2}\hbar^{-3/2}\int_{\Omega}b_p^{3/2}W_-d \mathbf{x}\}
\end{equation}

Suppose that $W_0$, $\mathbf{B}^0$, $Q=\cup_{k=1}^KQ_k $ satisfy ($A_1$)-($A_3$) of \S 2.
Then (\ref{Eqn3.2}) follows as before, and so do (\ref{Eqn3.7}) and (\ref{Eqn3.8}), with $W_0$
replaced by $W_0 +\lambda $. The remaining term on the right-hand
side of (\ref{Eqn3.7}) is estimated by (\ref{N46}). Since $(W_0+\lambda)_- \le (W_0)_-$,
we have, with $r=A\hbar$,
$$\begin{array}{l}
|\lambda| \hbar^3M_{\gamma}(\mathbb{P}_{W_0} + \lambda -C\hbar^2(
\rho r)^{-2}\chi_{\rho r}, S_{\rho r} \cap \Omega)\\
 \lsim  \int_{S_{\rho r} \cap \Omega}\big([W_0 -C\hbar^2(
\rho r)^{-2}\chi_{\rho r}]_-^{5/2} + b_p^{3/2}[W_0 -C\hbar^2(
\rho r)^{-2}\chi_{\rho r}]_-\big) d \mathbf{x}\\
 \lsim  \int_{S_{\rho r} \cap \Omega}\big([W_0]_-^{5/2} +
b_p^{3/2}[W_0]_-\big) d \mathbf{x}\\
\ \ \ + (\hbar/\rho r)^5|S_{\rho r}\cap Q_{\rho r}| +
(\hbar/\rho r)^2 \int_{S_{\rho r}\cap Q_{\rho r}} b_p^{3/2} d
\mathbf{x}\\
\lsim   K(\rho r) + (1/A\rho)^5(\rho +\rho r) +(1/A\rho)^2(\rho
+\rho r),\end{array}
$$
where $K(\rho r) \rightarrow 0 $ as $\rho \rightarrow 0$, uniformly
in $\hbar$ and $A$, by (\ref{Eqn3.5}). It follows as for (\ref{Eqn3.11}) that
\begin{equation}\label{N48}
\lim_{\hbar \rightarrow 0} \left\{ \hbar ^3M_{\gamma}(
\mathbb{P}_{W_0} +\lambda; \Omega)- \mathfrak{B}_{\gamma}(\mu
\hbar| \mathbf{B}|,W_0 +\lambda,\Omega)\right \} =0.
\end{equation}

For a general $W$ we proceed in a similar way to Sobolev in \cite{Sobolev1998},
using Shen's estimate (\ref{N46}).

Let $ \varepsilon >0 $ and choose non-overlapping cubes $Q_k$,
$k=1,..., K$, with sides parallel to the co-ordinate planes, and a
piecewise constant $W_0$ which is constant in each $Q_k$, is zero
outside $Q=\cup_{k=1}^K Q_k $ and
\begin{equation}\label{N49}
\int_{\Omega} |W-W_0|^{3/2} d \mathbf{x}< \varepsilon,
\int_{\Omega} b_p |W-W_0|^{1/2} d \mathbf{x}< \varepsilon.
\end{equation}
Note that assumptions 2. and 3. imply that $ W^{3/2}$, $b_pW^{1/2}
\in L^1(\Omega)$ and
\begin{equation}\label{N50}
\int_{\Omega} |W-W_0|^{5/2} d \mathbf{x} \lsim \varepsilon,\ \ \
\int_{\Omega} b_p^{3/2} |W-W_0| d \mathbf{x} \lsim \varepsilon.
\end{equation}
We prove the result for $\gamma =0$, the proof for $\gamma > 0$
being similar. Set $N_{\lambda}(W) \equiv N(\pw +\lambda, \Omega)$.
Then, we have the Weyl inequality
\[
N_{\lambda}(W) \ge N_{\lambda}(W_0/(1+\zeta))- N_{\lambda}([W-W_0]/\zeta)
\]
for any $ \zeta \in (0,1)$. By (\ref{N46}) and (\ref{N50})
\begin{equation}\label{N51}
\hbar^3N_{\lambda}([W-W_0]/\zeta) \lsim\  |\lambda|^{-1} \big
(\varepsilon \zeta^{-5/2} + \varepsilon \zeta^{-1}\big).
\end{equation}
Also, from (\ref{N48}),
\[
\lim_{\hbar \rightarrow 0}\left \{ \hbar^3N_{\lambda}(W_0/(1+\zeta))
- \mathfrak{B}_0(\mu \hbar| \mathbf{B}|, W_0/(1+
\zeta)+\lambda, \Omega)\right \} =0
\]
and from (\ref{Eqn2.10}) we have
\[
|\mathfrak{B}_0(\mu \hbar| \mathbf{B}|, W_0/(1+ \zeta) +\lambda,
\Omega)-\mathfrak{B}_0(\mu \hbar| \mathbf{B}|, W +\lambda,
\Omega)|
\lsim  \zeta ^{1/2} + \varepsilon ^{1/2}.
\]
Hence,
$$\begin{array}{l}
\liminf_{\hbar \rightarrow 0}\left \{\hbar^3
N_{\lambda}(W)-\mathfrak{B}_0(\mu \hbar| \mathbf{B}|, W+\lambda,
\Omega)\right \} \gsim \\
-|\lambda|^{-1}(\varepsilon
\zeta ^{-5/2}+ \varepsilon \zeta^{-1})- \zeta ^{1/2} -\varepsilon
^{1/2}.\end{array}
$$
Let $ \varepsilon \rightarrow 0$, $\zeta \rightarrow 0$ in that order
to get
\[
\liminf_{\hbar \rightarrow 0} \left \{\hbar^3
N_{\lambda}(W)-\mathfrak{B}_0(\mu \hbar| \mathbf{B}|, W+\lambda,
\Omega)\right \} \ge 0.
\]
The proof of the upper bound is similar. We use
\[
N_{\lambda}(W) \le \  N_{\lambda}(W_0/(1- \zeta))+
N_{\lambda}([W-W_0]/\zeta)
\]
and argue as before.
\end{proof}

The following result can be established similarly.

\begin{Theorem}\label{New3}
Suppose that
\begin{itemize}
\item [(i)] $W$, $\B$ are continuous,
\item [(ii)] $|W(\mathbf{x})|\to 0$ uniformly as
$|\mathbf{x}|\to\infty$ in $\Omega$, and
\item [(iii)] $\mu\hbar\le$ constant.
\end{itemize}
Then, (\ref{N45}) holds for all $\gamma\in [0,1)$ and $\lambda
>0$.
\end{Theorem}
\begin{proof}
The operator $\mathbb{P}_W(\Omega)$ is properly defined as a
self-adjoint operator on the domain of $\mathbb{P}_0(\Omega)$
since it is a bounded perturbation of $\mathbb{P}_0(\Omega)$. The
point to note is that $[W+\lambda]_-$ is compactly supported in
$\Omega$. On choosing $Q$ in ($A_1$) of \S2 to contain this
support, and the piecewise constant function $W_0$ to be such that
$W_\infty\equiv W-W_0\ge 0$ in $Q$, we have for
$$
\mathbb{P}_W+2\lambda = \{(1-\eta)\mathbb{P}_0+W_0+\lambda\} +
\{\eta\mathbb{P}_0+W_\infty+\lambda\}=: \mathbb{R}_1+\mathbb{R}_2
$$
that $\mathbb{R}_2\ge 0$ for all $\eta\in (0,1)$. The proof
follows easily from (\ref{N48}).
\end{proof}

\section{The Dirac operator: $\mu\hbar \le constant$}
Assume ($A_1$)-($A_3$) of \S2 with $W_0\equiv 0$. Set
$N(S,\Omega,I):=\#\{\lambda_n(S):\lambda_n(S)\in I\}$, the number
of eigenvalues of the operator $S(\Omega)$ in the interval $I$, and
$M_{\gamma}(S;\Omega,I)= \sum_{\lambda_n(S)\in I} |\lambda_n(S)|^{\gamma}$.
Then, with $V_0\ge 0$ in $Q_k$,
$$\begin{array}{l}
N(\dvo,Q_k,(-1,1)) + N(\mathbb{D}_{-V_0},Q_k,(-1,1))\\
= N(\dvo,Q_k,(-1,1))+N(\dvo, Q_k,(-1+2V_0,1+2V_0))\\
= N(\dvo,Q_k,(-1,1+2V_0)) + E,\end{array}
$$
where
$$
E:=\left \{\begin{array}{ll} N(\dvo,Q_k,(-1+2V_0,1))=0 & \text{if\ \
} V_0\le 1,\\
 -N(\dvo,Q_k,[1,-1+2V_0)) & \text{if\ \
} V_0> 1.\end{array}\right.
$$
For $\mathbb{R}_0:=\mathbb{D}_0^2-1= \mathbb{P}_0 \mathbb{I}_2$ and
$\lambda_{1}^o:=(1+V_0)^2-1$,
$$
N(\dvo,Q_k,(-1,1+2V_0)) = N(\mathbb{R}_0,Q_k,[-1,\lambda_{1}^o)),
$$
which implies that for any $\eta >0$
\begin{equation}\label{Eqn4.1}\begin{array}{l}
|N(\dvo,Q_k,(-1,1)) +
N(\mathbb{D}_{-V_0},Q_k,(-1,1))-N(\mathbb{R}_0,Q_k,[-1,\lambda_{1}^o))|\\
\le  N(\mathbb{R}_0,Q_k,[-1,\lambda_{-1}^o+\eta))\end{array}
\end{equation}
where $\lambda_{-1}^o:= (1-V_0)^2-1$. Inequality
(\ref{Eqn4.1}) holds as well for $V_0<0$ with
\begin{equation}\label{Eqn4.2} \lambda_{1}^o:=(1+|V_0|)^2-1,\q
\lambda_{-1}^o:=(1-|V_0|)^2-1. \end{equation}
For $\rvo\equiv \rvo(\B):=\dvo^2-1$, we have
$\rvo(\Omega)\le \oplus_{k=1}^K\rvo(Q_k)$. Since
\begin{equation}\label{Headache}
N(\dvo,\Omega,I)=N(\mathbb{R}_{V_0},\Omega)=
N(\mathbb{R}_{V_0},\Omega,(-\infty,0)), \q I=(-1,1),
\end{equation}
we may apply the minimax principle to $\mathbb{R}_{V_0}$ in order to
estimate $N(\dvo,\Omega,I)$. In particular,
$$\begin{array}{rl}
N(\dvo,\Omega,I)+N(\mathbb{D}_{-V_0},\Omega,I)\ge&
\sum_{k=1}^K\{N(\rvo,Q_k)+N(\mathbb{R}_{-V_0},Q_k)\}\\
=&\sum_{k=1}^K\{N(\dvo,Q_k,I)+N(\mathbb{D}_{-V_0},Q_k,I)\}. \end{array} $$
Hence, from (\ref{Eqn2.7}), (\ref{Eqn2.8}), and (\ref{Eqn4.1})
$$\begin{array}{l}
\frac12\hbar^3\{N(\dvo,\Omega,I)+N(\mathbb{D}_{-V_0},\Omega,I)\}\\ \ge
\frac12\hbar^3\sum_{k=1}^K\{N(\mathbb{R}_0,Q_k,[-1,\lambda_{1}^o))
-N(\mathbb{R}_0,Q_k,[-1,\lambda_{-1}^o+\eta))\}\\ \ge
(1-\delta)^3\mathfrak{B}(\mu\hbar|\B^o|,\frac{-\lambda_{1}^o}{1+\theta}
+\frac{(C\mu
r\sigma_r)^2}{\theta(1+\theta)}+\frac{C\hbar^2}{\delta^2r^2},Q)\\ \ \ \
-\mathfrak{B}(\mu\hbar|\B^o|,\frac{-\lambda_{-1}^o-\eta}{1-\theta}
-\frac{(C\mu r\sigma_r)^2}{\theta(1-\theta)},Q)\\
=(1-\delta)^3\mathfrak{B}(\mu\hbar|\B|,-\lambda_{1}^o,Q)
-\mathfrak{B}(\mu\hbar|\B|,-\lambda_{-1}^o,Q)\\ \ \ -O\Big (\big\{(\theta
+\eta +\frac{A^2\sigma_r^2}{\theta}+\frac{1}{A^2\delta^2})^\frac12
+(\theta +\eta
+\frac{A^2\sigma_r^2}{\theta}+\frac{1}{A^2\delta^2})^\frac32 +
\sigma_r^\frac14 +\sigma_r\big\}|Q|\Big )\end{array}
$$
for $r=A\hbar$, on
using Lemma~\ref{Eqn2.2} with $\mu\hbar\lsim 1$. It follows as for
(\ref{Eqn3.2}) that
\begin{equation}\label{Eqn4.3}\begin{array}{ll}
\liminf_{\hbar\to 0}&\big\{
\frac12\hbar^3[N(\dvo,\Omega,I)+N(\mathbb{D}_{-V_0},\Omega,I)]\\ &
-\mathfrak{B}(\mu\hbar|\B|,-\lambda_{1}^o,\Omega)
+\mathfrak{B}(\mu\hbar|\B|,-\lambda_{-1}^o,\Omega)\big\} \ge 0.\end{array}
\end{equation}
Note that $\lambda_{-1}^o\le 0$ if $|V_0|\le 2$, and consequently,
\begin{equation}\label{Eqn4.4}
\mathfrak{B}(\mu\hbar|\B|,-\lambda_{-1}^o,Q_k)=0\ \ \  \text{if\ \ }
|V_0|\le 2.
\end{equation}
To establish the reverse inequality in (\ref{Eqn4.3}), we first note that
for $\{\psi_k\}_{k=0}^K$ given in (\ref{Eqn3.4}) and all
$f\in[\Con(\Omega)]^4$
$$ \dvo^2[\psi_kf]=\psi_k\dvo^2f
+\frac{\hbar}{i}\sum_{j=1}^3\partial_j\psi_k[\alpha_j\dvo f+\dvo(\alpha_j
f)] -\hbar^2(\Delta \psi_k)f.
$$
Since $\sum_{k=0}^K\psi_k^2\equiv 1$, it
follows that $$\begin{array}{rl} \sum_{k=0}^K\psi_k\dvo^2[\psi_k f]=
&\dvo^2 f-\hbar^2\sum_{k=0}^K\psi_k(\Delta \psi_k)f\\ = &\dvo^2
f+\hbar^2\sum_{k=0}^K |\grad\psi_k|^2f.\end{array} $$
Therefore, by (\ref{Eqn3.4})
\begin{equation}\label{Eqn4.5}\begin{array}{rl}
(\mathbb{R}_{V_0}(\Omega)f,f)\ge &
\sum_{k=1}^K([\mathbb{R}_{V_0}(Q_k)-\frac{C\hbar^2}{\rho^2r^2}]\psi_k
f,\psi_k f)\\ &+ ([\mathbb{R}_{V_0}(S_{\rho
r}\cap\Omega)-\frac{C\hbar^2}{\rho^2r^2}\chi_{\rho r}]\psi_0 f,\psi_0
f).\end{array}
\end{equation}
Let $a^2:= C\hbar^2/(\rho r)^2$, where $C$
is the constant in (\ref{Eqn4.5}). Then,
$$
N(\mathbb{R}_{V_0}-a^2,Q_k)=N(\dvo,Q_k,(-\sqrt{1+a^2},\sqrt{1+a^2})),
$$
and on repeating the argument leading to (52), we have for any
$\eta >0$, with $\lambda_{\pm 1}^o(a)=(\sqrt{1+a^2}\pm |V_0|)^2
-1$,
\begin{equation}\label{Eqn4.6}\begin{array}{l}
\frac12\hbar^3\{N(\mathbb{R}_{V_0}-a^2,Q_k)+N(\mathbb{R}_{-V_0}-a^2,Q_k)\}
\\
\le
\frac12\hbar^3\{N(\mathbb{R}_0-\lambda_{1}^o(a),Q_k)+N(\mathbb{R}_0-
\lambda_{- 1}^o(a)-\eta,Q_k)\}\\ \le
\mathfrak{B}(\mu\hbar|\B^o|,-\frac{\lambda_{1}^o(a)}{1-\theta}-\frac{(C\mu
r\sigma_r)^2}{\theta(1-\theta)},Q_k)
+\mathfrak{B}(\mu\hbar|\B^o|,-\frac{\lambda_{-
1}^o(a)}{1-\theta}-\frac{\eta}{1-\theta}
-\frac{(C\mu r\sigma_r)^2}{\theta(1-\theta)},Q_k)\\
\le
\mathfrak{B}(\mu\hbar|\B|,-\lambda_{1}^o,Q_k)
+\mathfrak{B}(\mu\hbar|\B|,-\lambda_{-1}^o,Q_k)\\
+O\Big (\big\{(\theta +\eta +\frac{1}{A\rho}
+\frac{A^2\sigma_r^2}{\theta})^\frac12\\
\ \ \ \ \ \ \ \ +(\theta +\eta +\frac{1}{A\rho}
+\frac{A^2\sigma_r^2}{\theta}+)^\frac32
+ \sigma_r^\frac14 +\sigma_r\big\}|Q_k|\Big )
\end{array}
\end{equation}
by (\ref{Eqn2.8}) and (\ref{Eqn2.12}), since $r=A\hbar$. From Lemma~4.1 of \cite{EvansLewis1999}
\begin{equation}\label{Eqn4.7}\begin{array}{rl}
\hbar^3 N(\mathbb{R}_{V_0}-\frac{C\hbar^2}{\rho^2r^2}\chi_{\rho
r}, S_{\rho r}\cap \Omega)\lsim &
\int_{S_{\rho r}\cap \Omega}[|V_0|^\frac32 +|V_0|^3+|\B|^\frac32
+\frac{\hbar^3}{(\rho r)^3}\chi_{\rho r}]d\mathbf{x}\\
\lsim & \int_{\Omega\setminus Q}|\B(x)|^\frac32 d\mathbf{x} +
K(\rho r) +(A\rho)^{-3}[\rho + \rho r]
\end{array}
\end{equation}
 where
$$
K(\rho r):= \int_{S_{\rho r}\cap Q}[|V_0|^\frac32 + |V_0|^3
|\B|^\frac32]d\mathbf{x} \to 0, \q \text{as\ \ } \rho\to 0,
$$
uniformly in $\hbar$ and $A$ by (\ref{Eqn3.5}). As in (\ref{Eqn3.10}), on using
(\ref{Headache}), it follows from (\ref{Eqn4.5}),
(\ref{Eqn4.6}), and (\ref{Eqn4.7}) that
\begin{equation}\label{Eqn4.8}\begin{array}{rl} \limsup_{\hbar\to
0}&\big\{ \frac12\hbar^3[N(\dvo,\Omega,I)+N(\mathbb{D}_{-V_0},\Omega,I)]\\
& -\mathfrak{B}(\mu\hbar|\B|,-\lambda_{1}^o,\Omega)
-\mathfrak{B}(\mu\hbar|\B|,-\lambda_{-1}^o,\Omega)\big\} \le 0\end{array}
\end{equation} for $I=(-1,1)$.

Now, we are in a position to prove the first theorem in this
section.
\begin{Theorem}\label{Thm4.1}
Suppose that
\begin{itemize}
\item [(i)] $\B$ is continuous on $\Omega$,
\item [(ii)] $V,V^2,|\B|\in L^{\frac32+\gamma}(\Omega)$,
\item [(iii)] $|\{ \mathbf{x}\in \Omega: |V(\mathbf{x})|>2\}|=0$,
\item [(iv)] $\mu\hbar\le$ constant.
\end{itemize}
Then, with $I_\eta:=(-1+\eta,1-\eta)$, $\eta \in (0,1)$, and any
$\gamma \in[0,1]$,
$$
\lim_{\eta\to 0}\lim_{\hbar\to
0}\{\frac12\hbar^3[M_{\gamma}(\dv;\Omega,I_{\eta})+
M_{\gamma}(\mathbb{D}_{-V};\Omega,I_{\eta})]
- \mathfrak{B}_{\gamma}(\mu\hbar|\B|,-\lambda_1,\Omega)\}=0
$$
where $\lambda_1:= 2|V|+V^2$.
\end{Theorem}
\begin{proof}
We prove the result for $\gamma =0$, the other cases being
similar. As in the proof of Theorem~\ref{Count}, given $\epsilon >0$ choose
a piecewise constant function $V_0$ taking constant values in
non-overlapping congruent cubes $Q_k$, $k=1,\dots,K$, and such
that
\begin{equation}\label{Eqn4.9}
\|V-V_0\|_{\frac32,\Omega}<\epsilon,\ \ \ \|V-V_0\|_{3,\Omega}<\epsilon \
\ \ \|\B-\B^o\|_{\frac32,\Omega}<\epsilon
\end{equation}
in which $V_0\equiv 0$, $\B^o\equiv 0$ outside $Q:=\cup_{k=1}^KQ_k$. Thus, (\ref{Eqn4.1}) and
(\ref{Eqn4.8}) hold for $V_0$, and, indeed, for any function taking constant values in
each $Q_k$ and zero outside $Q$.
Let $\eta\in (0,1)$ and set
\begin{equation}\label{Eqn4.10}\begin{array}{rl}
\mathbb{D}_V=& (1-\eta)\{\al\cdot (\frac{\hbar}{i}\grad -\mu\sa)\}
+\beta +V +\eta\{\al\cdot(\frac{\hbar}{i}\grad -\mu\sa)\}\\ =&
\{\al\cdot (\frac{\hat h}{i}\grad -\hat\mu\sa)\} +\beta +V_0\} +
\{\eta\al\cdot(\frac{\hbar}{i}\grad -\mu\sa)+V_\infty\}\\ =:&
T_1+T_2\end{array}
\end{equation}
where $\hat h:=(1-\eta)\hbar$, $\hat\mu:=(1-\eta)\mu$, and $V_\infty
=V-V_0$. Then, for $\phi\in [\Con(\Omega)]^4$,
\begin{equation}\label{Eqn4.11}
([\mathbb{D}_V^2-1]\phi,\phi)=\|T_1\phi\|^2 +2\mathfrak{Re}(T_1\phi_1,
T_2\phi) +\|T_2\phi\|^2-\|\phi\|^2.
\end{equation}
With $\hat
D_\sa:=\frac{\hat h}{i}\grad -\hat\mu\sa$ and $D_\sa:=\frac{\hbar}{i}\grad
-\mu\sa$, we have
$$\begin{array}{rl} (T_1\phi,T_2\phi) =& (\{\al\cdot
\hat D_\sa +\beta+V_0\}\phi, \{\frac{\eta}{1-\eta}\al\cdot\hat D_\sa
+V_\infty\}\phi)\\ =&\frac{\eta}{1-\eta} \|T_1\phi\|^2 +
(T_1\phi,[V_\infty -\frac{\eta}{1-\eta} (\beta +V_0)]\phi).\end{array}
$$
Since $\mathfrak{Re}(\al\cdot\hat D_\sa\phi,\beta\phi)=0$,
$$\begin{array}{l} 2\mathfrak{Re}(T_1\phi,T_2\phi)=\\ \ \ \
\frac{2\eta}{1-\eta} \|T_1\phi\|^2 + 2\mathfrak{Re}[(T_1\phi,V_\infty\phi)
-\frac{2\eta}{1-\eta}(T_1\phi,V_0\phi)-\frac{2\eta}{1-\eta}((\beta+V_0)\phi,
\beta\phi)].\end{array}
$$
Hence,
\begin{equation}\label{Eqn4.12}\begin{array}{l}
|2\mathfrak{Re}(T_1\phi,T_2\phi)-
\frac{2\eta}{1-\eta} \|T_1\phi\|^2 +\frac{2\eta}{1-\eta}\|\phi\|^2|\\ \le
2\theta\|T_1\phi\|^2 +\frac{1}{\theta} (\|V_\infty\phi\|^2
+\frac{\eta^2}{(1-\eta)^2}\|V_0\phi\|^2)+\frac{2\eta}{1-\eta}
(|V_0|\phi,\phi).\end{array}
\end{equation}
Also,
$$
\|T_2\phi\|^2 =
\eta^2\|\al\cdot D_\sa\phi\|^2+ 2\eta \mathfrak{Re}(\al\cdot
D_\sa\phi,V_\infty\phi) +\|V_\infty\phi\|^2
$$
which implies that for any
$\theta >0$
\begin{equation}\label{Eqn4.13}\begin{array}{l} |\|T_2\phi\|^2
- \eta^2([H_0(\B)-\mu\hbar\sigma\cdot\B]\mathbb{I}_2\phi, \phi)|\\ \le
\theta\eta^2([H_0(\B)-\mu\hbar\sigma\cdot\B]\mathbb{I}_2\phi,
\phi) + (1+\frac{1}{\theta})|V_\infty\phi\|^2.\end{array}
\end{equation}
On substituting (\ref{Eqn4.12}) and (\ref{Eqn4.13}) in (\ref{Eqn4.11}), we
have that
\begin{equation}\label{N66}\begin{array}{rl}
([\mathbb{D}_V^2-1]\phi,\phi) \ge &
(1+\frac{2\eta}{1-\eta}-2\theta) \|T_1\phi\|^2
-\frac{1}{\theta}(\|V_\infty\phi\|^2+\frac{\eta^2}{(1-\eta)^2}\|V_0\phi\|^
2)\\
&-\frac{2\eta}{1-\eta}(|V_0|\phi,\phi)
+\eta^2(1-\theta)([H_0(\B)-\mu\hbar\sigma\cdot\B]\mathbb{I}_2\phi,\phi)\\
&-(1+\frac{1}{\theta})\|V_\infty\phi\|^2
-\|\phi\|^2-\frac{2\eta}{1-\eta}\|\phi\|^2\\
\ge &
(1+\frac{2\eta}{1-\eta}-2\theta)([\hat\mathbb{D}_{V_0}^2-1]\phi,\phi)
-\frac{\eta^2}{\theta(1-\eta)^2}\|V_0\phi\|^2\\
&-\frac{2\eta}{1-\eta}
(|V_0|\phi,\phi)
 -\eta^2(1-\theta)\mu\hbar(|\B^o|\phi,\phi)\\
&+\eta^2(1-\theta)([H_0(\B)-\mu\hbar|\B^\infty|]\mathbb{I}_2\phi,\phi)\\
&-(1+\frac{2}{\theta})(V_\infty^2\phi,\phi)
-2\theta\|\phi\|^2\end{array}
\end{equation}
 where $\hat\mathbb{D}_{V_0}=T_1=\al\cdot(\frac{\hat h}{i}\grad
-\hat\mu\sa) +\beta+V_0$. Whence, on choosing
$\theta=\eta/(1-\eta)<1/2$,
\begin{equation}\label{Eqn4.14}\begin{array}{rl}
\mathbb{D}_V^2-(1-\frac{2\eta}{1-\eta})\ge& \hat\mathbb{D}_{V_0}^2-1
-\frac{\eta}{1-\eta} (V_0^2+2|V_0|)-\eta^2\mu\hbar|\B^o|\\
&+\frac{\eta^2}{2}[H_0(\B)-\mu\hbar|\B^\infty|]\mathbb{I}_2
-\frac{2-\eta}{\eta}V_\infty^2\\ =:& S_1+S_2\end{array} \end{equation}
with
$$ S_1\equiv
S_1(V_0):=\hat\mathbb{D}_{V_0}^2-1 -\Phi,\q \Phi:= \frac{\eta}{1-\eta}
(V_0^2+2|V_0|)+\eta^2\mu\hbar|\B^o|,
$$
and
$$
S_2:=\frac{\eta^2}{2}[H_0(\B)-\mu\hbar|\B^\infty|
-\frac{2(2-\eta)}{\eta^3}V_\infty^2]\mathbb{I}_2.
$$
Since $\Phi$ is
piecewise constant and vanishes outside $Q$, the analysis leading to
inequality (58) holds with
$\mathbb{R}_{V_0}=\hat\mathbb{D}_{V_0}^2-1$ and $a^2=\Phi$, and this
yields the estimate
$$ \begin{array}{l}
\frac12\hat{h}^3\{N(S_1(V_0),\Omega)+N(S_1(-V_0),\Omega)\}\\
\le \mathfrak{B}(\hat\mu\hat{h}|\B|,-\lambda_{1}^o,\Omega)
+\mathfrak{B}(\hat\mu\hat{h}|\B|,-\lambda_{-1}^o,\Omega)\\
+O\big([\theta+\eta^{\frac12}+\frac{A^2\sigma_r^2}{\theta}]^{\frac12}
+[\theta+\eta^{\frac12}+\frac{A^2\sigma_r^2}{\theta}]^{\frac32}
+\sigma_r^\frac14 +\sigma_r \big)
\end{array}
$$ where $\lambda_{\pm
1}^o=(1\pm|V_0|)^2-1$, $\theta$ is arbitrary and $r=Ah$. Also, by (\ref{Eqn2.10})
$$ |\mathfrak{B}(\hat\mu\hat{h}|\B|,-\lambda_{\pm 1}^o,\Omega)
-\mathfrak{B}(\hat\mu\hat{h}|\B|,-\lambda_{\pm 1},\Omega)| = O(\|
V_\infty\|_{\frac32,\Omega}^\frac12 + \|V_\infty\|_{3,\Omega}
+ \| \mathbf{B}\|_{\frac32,\Omega}) \lsim \varepsilon^{1/2}
$$
with $\lambda_{\pm 1}:= (1\pm |V|)^2-1$.
 It follows that
\begin{equation}\label{Eqn4.15}\begin{array}{l}
\limsup_{\hbar\to 0}\big
\{\frac12\hat{h}^3 [N(S_1(V_0),\Omega)+N(S_1(-V_0),\Omega)]\\
\ \ \ \ \ -\mathfrak{B}(\hat\mu\hat{h}|\B|,-\lambda_1,\Omega)
-\mathfrak{B}(\hat\mu\hat{h}|\B|,-\lambda_{-1},\Omega)\big\}\lsim
\epsilon^\frac12 +\eta^\frac14.\end{array}
\end{equation}
A similar argument yields the reverse inequality (c.f.(\ref{Eqn4.3}))
\begin{equation}\label{Eqn4.16}
\begin{array}{l}
\liminf_{\hbar\to 0}\big
\{\frac12\hat{h}^3\{N(S_1(V_0),\Omega)+N(S_1(-V_0),\Omega)\}\\ \ \ \ \ \
-\mathfrak{B}(\hat\mu\hat{h}|\B|,-\lambda_1,\Omega)
+\mathfrak{B}(\hat\mu\hat{h}|\B|,-\lambda_{-1},\Omega)\big\}\gsim
-(\epsilon^\frac12 +\eta^\frac14).
\end{array}
\end{equation} Note that
$\lambda_{-1}=(1-|V|)^2-1\le 0$ for $|V|\le 2$, and also
$[2k\hat{\mu}\hat{h}|\B|-\lambda_{-1}]_-\le V^2$, and is zero for
$2k\hat{\mu}\hat{h}|\B|>\lambda_{-1}$. Therefore, we have
$$\mathfrak{B}(\hat{\mu}\hat{h}|\B|,-\lambda_{-1},\Omega)\lsim\int_{\Omega
\cap
\{x:|V_0(x)|>2\}}|V(x)|[|\B(x)|+|V(x)|]dx =0
$$
from assumption (iii).  Hence, as $\hbar \rightarrow 0$,
\begin{equation}\label{Eqn4.17}
\begin{array}{l}
\frac12\hat{h}^3\{N(S_1(V_0),\Omega)+N(S_1(-V_0),\Omega)\}
-\mathfrak{B}(\hat\mu\hat{h}|\B|,-\lambda_1,\Omega)\\
 = O(\epsilon^\frac12 +\eta^\frac12).\end{array}
\end{equation}
Finally, by \cite{EvansLewis1999}, Lemma~4.1,
\begin{equation}\label{Eqn4.18}
\hbar^3N(S_2,\Omega)\lsim\int_{\Omega}[|\B^\infty|^\frac32+\eta^{-\frac92}
V_\infty^3]d\mathbf{x}
\lsim \epsilon^\frac32+\epsilon^3\eta^{-\frac92}.
\end{equation}
It follows from (\ref{Eqn4.14}) ,(\ref{Eqn4.17}), and (\ref{Eqn4.18})
that, with $I_\eta\equiv (-1+\eta,1-\eta)$,
$$
\limsup_{\hbar\to
0}\{\frac12\hbar^3[N(\mathbb{D}_V,\Omega,I_\eta)+N(\mathbb{D}_{-V},\Omega,
I_\eta)]
-\mathfrak{B}(\mu\hbar|\B|,-\lambda_1,\Omega)\}\lsim\eta^\frac12.
$$
The reverse inequality is proved similarly, with $\eta<0$ in
(\ref{Eqn4.10}). The proof is complete.
\end{proof}
The next theorem is the analogue of Theorem~\ref{Count2} in \S3.
\begin{Theorem}\label{Thm4}
Suppose that
\begin{enumerate}
\item
$ \mathbf{B}$ is continuous,
\item
$| \mathbf{B}|$, $V \in L^{\infty}(\Omega)$,
\item
for some $p>3/2$, $V^5$, $b_p^{3/2} V^2 \in L^1(\Omega)$,
\item
$\mu \hbar \le$ constant,
\item
$|\{\mathbf{x}\in \Omega: |V(\mathbf{x})|>2\}| =0 $.
\end{enumerate}
Then, for all $\gamma \in [0,1]$,
\[
\lim_{\eta \rightarrow 0} \lim_{\hbar \rightarrow
0}\big\{\frac{\hbar^3}{2} \big[M_{\gamma}(\dv, \Omega,I_{\eta})+
M_{\gamma}( \mathbb{D}_{-V}, \Omega,I_{\eta})\big]-
\mathfrak{B}_{\gamma}(\mu \hbar| \mathbf{B}|,- \lambda_1, \Omega)
\big \} =0,
\]
where $I_{\eta}= (-1+\eta, 1-\eta), \eta \in (0,1)$ and $\lambda_1=2|V|+V^2$.
\end{Theorem}
\begin{proof}
The only difference with the proof of Theorem~\ref{Thm4.1} is in the way the
error terms involving $S_{\rho r}\cap \Omega $ are estimated to give
the analogue of (\ref{N48}). These are now dependent on (\ref{N46}) rather than on
the magnetic CLR inequality. We again prove it for $\gamma =0$.
From $(\dv-V)^2 = ( \mathbb{P}_0+1)\mathbb{I}_2$, it readily follows
that, for any $ \zeta >0$,
\[
\dv^2 \ge \frac{1}{1+ \zeta}[ \mathbb{P}_0 +1]-\frac{1}{\zeta}V^2
\]
and, for any $\lambda < 1$, on choosing $\zeta =
\lambda/(4-\lambda)$,
\[
\dv^2-1+\lambda \ge (3/4)\{ \mathbb{P}_0 +\lambda
-(4/\lambda)V^2\}.
\]
Consequently,
\[
 \mathbb{R}_{V_0} +\lambda -C\frac{\hbar^2}{(\rho r)^2}\chi_{\rho r}
 \ge (3/4)\left\{ \mathbb{P}_0+\lambda
 -(4/\lambda)V_0^2-C\frac{\hbar^2}{(\rho r)^2}\chi_{\rho
 r}\right\}
 \]
 and, from (\ref{N46}), with $r=A\hbar $,
$$\begin{array}{l}
 \hbar^3  N( \mathbb{R}_{V_0}+\lambda -C\frac{\hbar^2}{(\rho r)^2}\chi_{\rho
 r}, S_{\rho r}\cap \Omega)\\
 \lsim \int _{S_{\rho r}\cap
 \Omega}\big\{[V_0^2+(1/A\rho)^2\chi_{\rho
 r}]^{5/2}+b_p^{3/2}[V_0^2+1/(A\rho)^2\chi_{\rho r}]\}d\mathbf{x}\\
  \lsim  K(\rho r)+(1/A\rho)^5(\rho+\rho r)+(1/A
 \rho)^2(\rho+\rho r),\end{array}
$$
as in the proof of Theorem~\ref{Count2}, where $K(\rho r) \rightarrow 0 $ as
$\rho \rightarrow 0$, uniformly in $\hbar $ and $A$. This gives
(\ref{Eqn4.8}) as in Theorem~\ref{Count2}.
 From (\ref{N66}) we have,
\[
 \dv^2-1+2\lambda+2\theta \ge (S_1 +\lambda)+(S_3+\lambda),
\]
where, with $\theta=\eta /(1-\eta )$,
\[
S_3=\eta(1-2\eta) \mathbb{P}_0( \mathbf{B}^{\infty})
-(1/\theta)V_{\infty}^2.
\]
The $S_1+\lambda$ term is handled as in the proof of Theorem~\ref{Count2} and the
first part above. The $S_3$ term on $S_{\rho r}\cap\Omega $ is again
estimated by (\ref{N46}).
\end{proof}

\section{The Pauli Operator: magnetic fields with constant
direction.}
First, we summarize some facts from \cite{Sobolev1998} that we
will need below. Let $\B( \mathbf{x})=(0,0,B(x_1,x_2))$,
$ \mathbf{x}= (x,x_3), x=(x_1,x_2)$.
Assume that $B=B(x_1,x_2)$ is periodic, i.e., for some $T_1,T_2$
$$
B(x_1,x_2)=B(x_1+T_1,x_2)=B(x_1,x_2+T_2).
$$
For $T_3>0$ denote by $D^{(3)}=\times_{k=1}^3[0,T_k)$ the
fundamental domain of the lattice $\Gamma$ with vertices
$(T_1m_1,T_2m_2,T_3m_3)$, $m_j\in\mathbb{Z}$, $j=1,2,3$. Choose
the following vector potential $\sa$ corresponding
to $\B$: for $\phi=\phi(x_1,x_2)$
$$
\sa(x)=(-\partial_2\phi,\partial_1\phi,0),\q \Delta\phi
=B.
$$
Set
$$\begin{array}{rl}
B_0=&\frac{1}{T_1T_2}\int_{0}^{T_1}\int_0^{T_2}B(x_1,x_2)dx_1dx_2,\q
\B_0:=(0,0,B_0),\\
B_1=& B-B_0,\\
\phi_0=& \frac{B_0}{4}(\alpha x_1^2+\beta x_2^2),\q \alpha +\beta
=2,\\
\phi = & \phi_0+\phi_1,\end{array}
$$
where $\Delta \phi_1 =B_1$ and $\phi_1$ is periodic on
$D^{(2)}=[0,T_1)\times [0,T_2)$. Note that the conditions above
imply that $B_1$ is periodic,
$\int_0^{T_1}\int_0^{T_2}B_1dx_1dx_2=0$, and $\Delta\phi_0=B_0$.
The {\it flux} of $B$ across $D^{(2)}$ is
$$
\Phi = \frac{1}{2\pi}\int_{D^{(2)}}B(x_1,x_2)dx_1dx_2.
$$
The physics of our system must be invariant under translations
$(x_1, x_2)\rightarrow (x_1 +T_1, x_2), (x_1, x_2) \rightarrow (x_1,
x_2+T_2)$, since the magnetic field is invariant under these
translations. The magnetic potential $ \mathbf{a}$ is not
invariant, but since $ \mathbf{a}$ and the translated $ \mathbf{a},
{\mathbf{a}}_{T_j}, j=1,2, $ say, give rise to the same magnetic field, there
must be gauge transformations
$$
(\tau_ju)(\mathbf{x}):= e^{if_j(\mathbf{x})}u_{T_j}(\mathbf{x}),\q j=1,2,
$$
in which
$$
u_{T_1}(\mathbf{x})= u(x_1+T_1,x_2),\q u_{T_2}(\mathbf{x})= u(x_1,x_2+T_2),
$$
which are unitary maps on the Hilbert space and
$$
(\frac{\hbar}{i}\grad -\mu\sa)(\tau_j u) =
e^{if_j}(\frac{\hbar}{i}\grad -\mu\sa_{T_j})u_{T_j},\q j=1,2.
$$
From the discussions above, we see that
$$
\sa =\sa^0+\sa^1,\q
\sa^k:=(-\partial_2\phi_k,\partial_1\phi_k,0),\q k=0,1,
$$
where $\sa^1$ is periodic and
$$\begin{array}{rl}
\sa^0(x_1+T_1,x_2) =& \sa^0(x_1,x_2) + \frac{B_0}{2}\alpha
T_1(0,1,0)\\
\sa^0(x_1,x_2+T_2) =& \sa^0(x_1,x_2) - \frac{B_0}{2}\beta
T_2(1,0,0).\end{array}
$$
The choice $f_1(\mathbf{x})=-i\frac{\mu\alpha B_0T_1}{2\hbar}x_2$ and
$f_2(\mathbf{x})=i\frac{\mu\beta B_0T_2}{2\hbar}x_1$ yield the
 ``magnetic translations" given by
$$\begin{array}{rl}
(\tau_1u)(x_1,x_2,x_3)=& u(x_1+T_1,x_2,x_3)\exp(-i\mu\alpha
B_0T_1x_2/2\hbar),\\
(\tau_2u)(x_1,x_2,x_3)=& u(x_1,x_2+T_2,x_3)\exp(i\mu\beta
B_0T_2x_1/2\hbar).\end{array}
$$
They commute with the expressions
$\Pi_k=-i\hbar\partial_k-\mu a_k$, $k=1,2,3,$ and $Q_\pm:=
\Pi_1\pm i\Pi_2$. If the flux of $ \mathbf{B}$ satisfies the condition
\begin{equation}\label{N73}
\mu\hbar^{-1}\Phi=N\in\mathbb{Z}
\end{equation}
then $\tau_1 $ and $\tau_2$ commute and we can reduce our problem
to one on the torus $X^{(3)} = \mathbb{R}^3/\Gamma$. In this case, we define the
operators $\Pi_k=-i\hbar\partial_k-\mu a_k$, and $Q_{\pm}$ in
$L^2(X^{(3)})$ with domains consisting of functions $u\in C^\infty(\R^3)$ which
satisfy
$$
\tau_ku=u,\ \ k=1,2,\q u(x_1,x_2,x_3+T_3)=u(x_1,x_2,x_3).
$$
and denote them by $Q_\pm(X^{(3)})$, $\Pi_k(X^{(3)})$. It is proved
in \cite{Sobolev1998} that these operators are closable, that each $\Pi_k $ is
symmetric and $Q_{\pm}^*\subset Q_{\mp} $. Let
$$\begin{array}{rl}
A_\pm(X^{(3)}):=& H_0(X^{(3)})\mp\mu\hbar B\\
\equiv & Q_\pm^*(X^{(3)})Q_\pm(X^{(3)}) +
\Pi_3^2(X^{(3)})\end{array}
$$
where $H_0=H_0(\B)$, $\B=\grad\times\sa$, as defined in \S1. We denote the closures of
the operators by the same notation.
The Pauli operator on the torus $X^{(3)}$, $\mathbb{P}(X^{(3)})$, is now
defined by
\[
\mathbb{P}_0(X^{(3)})=\left (\begin{array}{cc} A_+(X^{(3)}) &
0\\ 0 & A_-(X^{(3)})\end{array}\right )
\]
and $\pw(X^{(3)}) = \mathbb{P}_0(X^{(3)})+W $. We remind the reader that
$\pw(\Omega), \pw(Q) $ will always stand for the operator $\pw$
with Dirichlet boundary conditions.

Using a result\footnote{See
also Appendix A of Sobolev~\cite{Sobolev1998} for a proof.}
due to Dubrovin and Novikov~\cite{DubrovinNovikov1980}
that if $\pm N>0$ ($N$ as defined in (\ref{N73})) $\lambda=0$ is an eigenvalue of $A_\pm(X^{(3)})$ of multiplicity
$\pm N$, Sobolev proves
\begin{Lemma}\label{Lem5.1} (Sobolev~\cite{Sobolev1998},
Lemma~4.3, $\gamma=0$, $d=3$) Let (\ref{N73}) be satisfied and $|B|\ge
\kappa>0$. If $W$ is constant in $D^{(3)}$ and $W_-<2\mu\hbar\kappa$, then
$$ \big | N(\mathbb{P}_W,X^{(3)})-\frac{T_3}{\pi\hbar}W_-^\frac12|N|\big
|\le |N|. $$ \end{Lemma}
We shall follow Sobolev's strategy, which is again basically inspired by the
method of Colin de Verdiere~\cite{ColindeVerdiere1986}, but substantially
modified to meet the needs of the constant direction magnetic field case, using the
observations and results noted above. There will now be an infinity of
negative eigenvalues in general, and so we can only expect results for
$N(\pw+\lambda,\Omega)$ with $\lambda >0$; in fact for
$M_{\gamma}(\pw+\lambda,\Omega)$, $\lambda >0 $ when $\gamma \in [0,1/2]$.
When $\gamma >1/2$, $M_{\gamma}(\pw,\Omega) $ is finite
and its semi-classical limit is given by Sobolev in \cite{Sobolev1998}. Our analysis
requires an estimate for $\hbar^3N(\pw+\lambda,\Omega), \lambda >0,$ in
which $\mu |B|$ occurs linearly. A suitable result is obtained by Shen in
\cite{Shen1999}. In it $ N(\pw+\lambda, \Omega)$ grows like $\lambda^{-1/2}$ as
$ \lambda \rightarrow 0+$, and we are forced to take $\lambda > 0$
in our result for $ M_{\gamma}(\pw+\lambda,\Omega)$ when $ \gamma \le 1/2$ in view of (\ref{N26}).
If $\Omega$ is a bounded domain we prove in the next lemma that the $\lambda$
dependence in $N(\pw+\lambda,\Omega)$ is at most logarithmic; of course,
this is still crude since the spectrum of $\pw(\Omega)$ is discrete in this
case, but it does have interesting implications for $M_{\gamma}(\pw,\Omega)$
for $\gamma >0$, and, moreover, the technique used to prove it can be made
to yield a similar estimate for unbounded domains; see Proposition~\ref{Prop5} below.
\begin{Lemma}\label{Thm5.2} Let
$Q=Q^{(2)}\times [0,R)$, $W_-\in L^\infty(Q)$, and $\B=(0,0,B)$ .
Then, for $\lambda >0$ and $\tilde W:=\inf_{Q}W(x)$
\begin{equation}\label{Eqn5.2}
N(\mathbb{P}_W+\lambda,Q)\lsim [(\tilde
W_-+1)e^{\tilde W_-}|\log\lambda^{-1}|]\mu\hbar^{-2}(\tilde
W+\lambda)_-^\frac12 |Q|\max_{Q^{(2)}}|B(x)|. \end{equation}
\end{Lemma}
\begin{proof} We have that $$
\mathbb{P}_W=\mathbb{P}_0^{(2)}+W-\hbar^2\partial_3^2\ge
\mathbb{P}_0^{(2)}+\tilde W -\hbar^2\partial_3^2 = \mathbb{P}_{\tilde
W} $$ where $\mathbb{P}_0^{(2)}$ is the Pauli operator in two
dimensions, namely
$$
\mathbb{P}_0^{(2)}=\left (\begin{array}{cc} A_+^{(2)} &
0\\ 0 & A_-^{(2)}\end{array}\right )
$$
with
$$
A_\pm^{(2)} = Q_\pm^*Q_\pm =
H_0^{(2)}\mp\mu\hbar B
$$
and
$$
H_0^{(2)}=\sum_{k=1}^2(\frac{\hbar}{i}\partial_k-\mu a_k)^2
$$
the magnetic Schr\"odinger operator in $\R^2$. The eigenvalues of
$-\hbar^2\partial_3^2$ with Dirichlet boundary conditions on $(0,R)$ are
$\epsilon_m=(m\pi\hbar)^2/R^2$, $m\in\mathbb{N}$. It follows that the
eigenvalues of $\mathbb{P}_{\tilde W}$ are of the form
$\lambda_n(\mathbb{P}_0^{(2)}+\tilde W )+\epsilon_m$, where
$\lambda_n(\mathbb{P}_0^{(2)}+\tilde W )$ are the eigenvalues of
$\mathbb{P}_0^{(2)}+\tilde W $ on $Q^{(2)}$ (with Dirichlet boundary
conditions). Hence, for any $\lambda\in\R$,
\begin{equation}\label{Eqn5.3}\begin{array}{rl}
N(\mathbb{P}_W+\lambda,Q)\le & \sum_{m}N(\mathbb{P}_0^{(2)}+\tilde W
+\epsilon_m+\lambda,Q^{(2)})\\ =&
\sum_{m}\sum_{j=1}^2N(H_0^{(2)}+(-1)^j\mu\hbar B+\tilde
W +\epsilon_m+\lambda,Q^{(2)}).
\end{array}
\end{equation}
Since $\mathbb{P}_0^{(2)}\ge 0$, we have $A_\pm^{(2)}=H_0^{(2)}\mp\mu\hbar B\ge 0$.
Hence, for any $\epsilon \in (0,1)$
$$\begin{array}{rl}
H_0^{(2)}\mp\mu\hbar B+\tilde W +\epsilon_m+\lambda \ge &
\epsilon(H_0^{(2)}\mp\mu\hbar B)+\tilde W +\epsilon_m+\lambda\\ \ge &
\epsilon H_0^{(2)} -\epsilon\mu\hbar |B|+ \tilde
W +\epsilon_m+\lambda.
\end{array}
$$
Replacing $A_\pm $ by something smaller seems rather wasteful, but will be
seen to provide the correct scaling for $\mu|B|$ by a suitable choice of
$ \varepsilon $. It is reminiscent of a technique (running energy-scale
renormalization) used in a paper of Lieb, Loss and Solovej~\cite{Liebetal1995A}.
On substituting in (\ref{Eqn5.3}),
this yields
\begin{equation}\label{Eqn5.4}
N(\mathbb{P}_W+\lambda,Q)\le
2\sum_m N(\epsilon H_0^{(2)}-\epsilon\mu\hbar|B|+ \tilde
W +\epsilon_m+\lambda,Q^{(2)})
\end{equation}
where the sum is over all $m$ such that
\begin{equation}\label{Eqn5.5} \epsilon_m  =
\frac{(m\pi\hbar)^2}{R^2} < (\tilde W +\lambda)_-.
\end{equation}
To estimate $N(\epsilon
H_0^{(2)}-\epsilon \mu \hbar|B|-\tilde
W_-+\epsilon_m+\lambda,Q^{(2)})$ we use the result of Rozenblum
and Solomyak~[\cite{RozenblumSolomyak1997}, Theorem~2.4] with
(in their notation)
$$
\mathcal{A}=\epsilon H_0^{(2)}+\epsilon_m+\lambda,\ \
\mathcal{B}=-\epsilon\hbar^2\Delta+\epsilon_m+\lambda,\ \
V=(\epsilon\mu\hbar|B|+\tilde W_-)\chi_{Q^{(2)}}.
$$
Then, $\mathcal{B}\in \mathcal{P}$ as it generates a positivity
preserving contractive semigroup with the $(2,\infty)$-boundedness
(ultracontractivity) property having kernel $Q(t,x,y)$, where, on
the diagonal,
$$
Q(t,x,x)= \frac{1}{2\pi\hbar^2\epsilon
t}e^{-(\epsilon_m+\lambda)t};
$$
see \cite{RozenblumSolomyak1997}, \S2, for the terminology. The operator $\mathcal{A}$ is dominated by
$\mathcal{B}$ in the sense that
$$
|e^{-t\mathcal{A}}\psi(x)|\le e^{-t\mathcal{B}}|\psi(x)|\ \ \ \text{for a.e.\ }
x\in\R^2,\ \ \ x=(x_1,x_2),
$$
i.e., $\mathcal{A}\in \mathcal{P}\mathcal{D}(\mathcal{B})$ in the language of
\cite{RozenblumSolomyak1997}, \S2.3.
It follows from [\cite{RozenblumSolomyak1997}, Theorem~2.4] with
the choice $G(z)=(z-k)_+$ for some $k>0$, that
\begin{equation}\label{Eqn5.6}\begin{array}{rl}
N(\mathcal{A}-V)\le& \frac{1}{g(1)}\int_0^\infty \frac{dt}{t}\int_{Q^{(2)}}
\frac{1}{2\pi\hbar^2\epsilon
t}e^{-(\epsilon_m+\lambda)t}(tV(x)-k)_+dx\\
=& \frac{1}{2\pi g(1)}\int_{Q^{(2)}}dx\int_{k/V(x)}^\infty
e^{-(\epsilon_m+\lambda)t}(tV(x)-k)\frac{dt}{\epsilon\hbar^2t^2}\\
=& \frac{1}{2\pi\hbar^2\epsilon
g(1)}\int_{Q^{(2)}}V(x) dx \int_{\phi(x)}^\infty
e^{-(\epsilon_m+\lambda)t}(t-\phi(x))\frac{dt}{t^2}
\end{array}
\end{equation}
for $\phi(x):= k/V(x)$ and
\begin{equation}\label{Eqn5.7}
g(1)=\int_0^\infty(z-k)_+e^{-z}\frac{dz}{z}=\int_1^\infty
e^{-ks}\frac{ds}{s^2}.
\end{equation}
Now choose $\epsilon =\tilde W_-/(\mu\hbar\max_{Q^{(2)}}|B(x)|)$,
$k=2\tilde W_-$. Then $\phi(x)\ge 1$ and
$$
\int_{\phi(x)}^\infty
e^{-(\epsilon_m+\lambda)t}(t-\phi(x))\frac{dt}{t^2}\le \int_{1}^\infty
e^{-\lambda t}(t-1)\frac{dt}{t^2} \le \int_{1}^\infty e^{-\lambda
t}\frac{dt}{t}\le |\log \lambda^{-1}| + O(1)
$$
as $\lambda\to 0$.
Furthermore, on integration by parts,
$$
g(1)=\frac{e^{-k}}{k}-\frac{2}{k}\int_1^\infty e^{-ks}\frac{ds}{s^3}\ge
\frac{e^{-k}}{k}-\frac{2}{k}g(1)
$$
implying that
\begin{equation}\label{Eqn5.8} g(1)\ge e^{-k}/(k+2).
\end{equation} It now
follows from (\ref{Eqn5.4}), (\ref{Eqn5.5}), (\ref{Eqn5.6}), and
(\ref{Eqn5.8}) that
$$ N(\mathbb{P}_W+\lambda,Q) \lsim (\tilde W
+\lambda)_-^\frac12 \frac{R}{\pi\hbar}(k+2)e^k(|\log
\lambda^{-1}|)\mu\hbar^{-1}|Q^{(2)}|\max_{Q^{(2)}}|B(x)|
$$ and
(\ref{Eqn5.2}) is proved.
\end{proof}

The method of proving Lemma~\ref{Thm5.2} also yields
Proposition~\ref{Prop5} at the end of this section for an
unbounded $\Omega$, and an operator $\mathbb{P}_W(\Omega)$ with a
non-empty essential spectrum. However, we leave the result till
then so as not to break the flow of the argument leading to the
proof of Theorem~\ref{Thm5.6}.

\begin{Lemma}\label{Lem5.3} Let $Q=[0,R)^3$, $W$  constant on
$Q$, and $|B|\ge \kappa >0$. Then, for any $\lambda >0$,
\begin{equation}\label{Eqn5.9}
\lim_{\hbar\to 0}\{\mu^{-1}\hbar^2
N(\mathbb{P}_W+\lambda,Q)-\frac{1}{2\pi^2}\int_Q|B|(W+\lambda)_-^\frac12
d\mathbf{x}\}=0
\end{equation}
with $2\mu\hbar\kappa \ge W_-$. For any $\gamma >0$
\begin{equation}\label{Eqn5.10}
\lim_{\hbar\to 0} \{\mu^{-1}\hbar^2
M_\gamma(\mathbb{P}_W,Q)-\beta_\gamma\int_Q|B|W_-^{\gamma
+\frac12}d\mathbf{x}\}=0
\end{equation}
with $2\mu\hbar\kappa \ge W_-$, in which
$$
\beta_\gamma :=\frac{1}{4\pi^2}\int_0^1 t^{\gamma}(1-t)^{-\frac12} dt.
$$
\end{Lemma}
\begin{proof}
As in [\cite{Sobolev1998}, Lemma~5.1], let $\Omega_j^{(2)}$, $j\in
\mathbb{N}$, be non-overlapping squares obtained by translating
$Q^{(2)}=[0,R)^2$ parallel to the coordinate axes to form a
tesselation of $\R^2$. Choose a square $D^{(2)}\subset \R^2$ such
that
$$
\mu\hbar^{-1}\frac{1}{2\pi}\int_{D^{(2)}}B(x)dx = N\in
\mathbb{Z}
$$
with $B$ extended by periodicity from $Q^{(2)}$ to $\R^2$. Let
$M=\#\{j:\Omega_j^{(2)}\subset D^{(2)}\}$ and suppose, without
loss of generality, that
\begin{equation}\label{Eqn5.11}
(1-\epsilon)|N|\le \mu\hbar^{-1}M|\Phi|\le |N|,\q \Phi
=\frac{1}{2\pi}\int_{Q^{(2)}}B(x)dx
\end{equation}
for $\epsilon\in (0,\frac12)$. Set $D^{(3)}=D^{(2)}\times [0,R)$
and denote by $X^{(d)}$ the torus with fundamental domain
$D^{(d)}$, $d=2,3$.
It is proved in [\cite{Sobolev1998}, (5.3)] that if
$[W+\lambda]_-<2\mu\hbar\kappa$
\begin{equation}\label{Eqn5.12}
\limsup_{\hbar\to 0}\mu^{-1}\hbar^2 N(\mathbb{P}_W+\lambda,Q)\le
\frac{1}{2\pi^2}\int_Q|B|(W+\lambda)_-^\frac12 d \mathbf{x},\q \forall
\lambda\ge 0.
\end{equation}
For the lower bound we need $\lambda >0$ in order to use
Lemma~\ref{Thm5.2}. Let
$$\begin{array}{rl}
S^{(2)} =&\overline{X^{(2)}}\setminus \cup_{j=1}^M int(\Omega_j^{(2)})\\
S_\delta^{(2)}=& \{x\in X^{(3)}:\ dist(x,S^{(2)})<\delta\}
\end{array}
$$ for
$\delta\in (0,\frac{R}{2})$, and set
$S^{(3)}=S^{(2)}\times [0,R)$,
$S_\delta^{(3)}=S_\delta^{(2)}\times [0,R)$. Then, $X^{(d)}\subseteq
S_\delta^{(d)} \cup_{j=1}^M int(\Omega_j^{(d)})$ for $d=2,3$. Also,
since $|B|\ge \kappa$
$$
\begin{array}{rl} \frac{1}{2\pi}\kappa
|S^{(2)}|\le& \frac{1}{2\pi}\int_{S^{(2)}}|B|dx\\ =&
\frac{1}{2\pi}\int_{D^{(2)}}|B|dx
-\frac{1}{2\pi}M\int_{\Omega_j^{(2)}}|B|dx\\ \le& \frac{\epsilon
|N|}{\mu\hbar^{-1}}
\end{array}
$$
from (\ref{Eqn5.11}); note that $B$ is of one sign since $B$ is continuous
and $|B|\ge \kappa$. Therefore,
$$
|S^{(2)}|\le \frac{2\pi}{\kappa\mu\hbar^{-1}}\epsilon |N|
$$
and
$$
|S_\delta^{(2)}|\lsim \frac{\mu^{-1}\hbar\epsilon}{\kappa} |N| +M\delta
$$
which implies that
\begin{equation}\label{Eqn5.13}\begin{array}{rl}
|S_\delta^{(3)}|\lsim &  \frac{\mu^{-1}\hbar\epsilon}{\kappa} |N|R
+MR\delta\\ \le& \frac{M\epsilon}{\kappa(1-\epsilon)}|\Phi|R+MR\delta\\
\le& C(\epsilon +\delta)MR\end{array}
\end{equation} where $C$ depends on
$\kappa$ and $\Phi$.  We now proceed as in (\ref{Eqn3.6}). The analogue is
$$\begin{array}{rl}
([\mathbb{P}_{W+\lambda}(X^{(3)})+\frac{C\hbar^2}{\delta^2}]f,f)\ge &
(\mathbb{P}_{W+\lambda}({S}_{\delta}^{(3)})\psi_0f,\psi_0f)\\
&+\sum_{j=1}^M(\mathbb{P}_{W+\lambda}(\Omega_j^{(3)})\psi_j f,\psi_jf)
\end{array}
$$ for $f$ in the domain of $\mathbb{P}_W(X^{(3)})$ and a
partition of unity $\{\psi_j\}_0^M$ subordinate to the covering of
$X^{(3)}$ by $S_\delta^{(3)} \cup_{j=1}^M\Omega_j^{(3)}$ which
satisfies (\ref{Eqn3.4}) with $K=M$ and $\delta$ replacing $\rho r$. From
this we conclude that
\begin{equation}\label{Eqn5.14}
\begin{array}{rl}
N(\mathbb{P}_{W}+\lambda +\frac{C\hbar^2}{\delta^2},X^{(3)})\le
N(\mathbb{P}_{W}+\lambda, {S}_{\delta}^{(3)}) + M
N(\mathbb{P}_{W}+\lambda,Q)
\end{array}
\end{equation}
From Lemma~\ref{Lem5.1}
\begin{equation}\label{Eqn5.15}
N(\mathbb{P}_{W}+\lambda +\frac{C\hbar^2}{\delta^2},X^{(3)}) \ge
\frac{1}{\pi\hbar}R(W+\lambda+\frac{C\hbar^2}{\delta^2})_-^\frac12 |N|-|N|
\end{equation}
and by Lemma~\ref{Thm5.2} and (\ref{Eqn5.13})
\begin{equation}\label{Eqn5.16} N(\mathbb{P}_{W}+\lambda,
{S}_{\delta}^{(3)})\lsim
\mu\hbar^{-2}|\log\lambda^{-1}||S_\delta^{(3)}|\lsim
\mu\hbar^{-2}|\log\lambda^{-1}|(\epsilon +\delta)MR.
\end{equation} It
follows from (\ref{Eqn5.14})-(\ref{Eqn5.16}) that
$$
MN(\mathbb{P}_W+\lambda,Q)\ge\frac{1}{\pi\hbar}R(W+\lambda+\frac{C\hbar^2}
{\delta^2})_-^\frac12
|N|-|N|-C\mu\hbar^{-2}|\log\lambda^{-1}|(\epsilon +\delta)RM.
$$
Since
$$
|(W+\lambda+\frac{C\hbar^2}{\delta^2})_-^\frac12 -
(W+\lambda)_-^\frac12|\lsim \hbar/\delta
$$
we have from (\ref{Eqn5.11})
$$
\mu^{-1}\hbar^2 N(\mathbb{P}_W+\lambda,Q)\ge
\frac{R}{\pi}(W+\lambda)_-^\frac12 |\Phi|
-C(\frac{\hbar}{\delta}R|\Phi|+\hbar|\Phi|+|\log\lambda^{-1}|(\epsilon
+\delta)R)
$$
which yields (\ref{Eqn5.9}) on recalling (\ref{Eqn5.12}).
Note that (\ref{Eqn5.16}) gives,
for any $\gamma >0$,
\begin{equation}\label{Eqn5.17}
M_\gamma(\mathbb{P}_W,S_\delta^{(3)}) = \gamma \int_0^\infty
\lambda^{\gamma -1} N(\mathbb{P}_W+\lambda,S_\delta^{(3)})d\lambda\lsim
\mu\hbar^{-2}(\epsilon +\delta)RM.
\end{equation}
A similar argument then gives (\ref{Eqn5.10}).
\end{proof}
For $|B|\ge \kappa$ and $2\mu\hbar \kappa \ge W_-$ we have
$2k\mu\hbar|B|+W\ge 2\mu\hbar\kappa-W_-\ge 0$, for $k\ge 1$, and, for any $\gamma\ge 0$,
$$
\mathfrak{B}_\gamma(\mu\hbar|\B|,W,Q)=\mu\hbar
\beta_\gamma\int_Q|B|W_-^{\gamma+\frac12}d\mathbf{x}.
$$ This fact gives
\begin{Corollary}\label{Cor5.4} Let $Q=[0,R)^3$, $W$ a constant on $Q$,
and $|B|\ge \kappa >0$. Then, for any $\lambda >0$,
\begin{equation}\label{Eqn5.18} \lim_{\hbar\to 0}\frac{\hbar^3}{\mu\hbar
+1}\{
N(\mathbb{P}_W+\lambda,Q)-\hbar^{-3}\mathfrak{B}_0(\mu\hbar|\B|,W+\lambda,
Q)\}=0
\end{equation}
uniformly in $\mu\ge 0$. Moreover, for any $\gamma >0$
$$
\lim_{\hbar\to 0} \frac{\hbar^3}{\mu\hbar +1}\{M_\gamma(\mathbb{P}_W,Q)
-\hbar^{-3}\mathfrak{B}_\gamma(\mu\hbar|\B|,W,Q)\}=0.
$$
\end{Corollary}
In order to remove the assumption that $|B|\ge \kappa$, we prove
the following
\begin{Lemma}\label{Lem5.5} Let $Q=[0,R)^3$ and $W$ a nonzero constant on
$Q$. Then, for $\lambda >0$
\begin{equation}\label{Eqn5.19}\begin{array}{rl}
\limsup_{\hbar\to 0}\frac{\hbar^3}{\mu\hbar
+1}&|N(\mathbb{P}_W+\lambda,Q)-\hbar^{-3}\mathfrak{B}_0(\mu\hbar|\B|,W+
\lambda,Q)|\\
\le & C|\log\lambda^{-1}| W_-^\frac12
|Q|\max_{Q^{(2)}}|B(x)|,\end{array}
\end{equation}
and for any $\gamma >0$
\begin{equation}\label{Eqn5.20}\begin{array}{rl}
\limsup_{\hbar\to 0}\frac{\hbar^3}{\mu\hbar
+1}&|M_\gamma(\mathbb{P}_W,Q)-\hbar^{-3}\mathfrak{B}_\gamma(\mu\hbar|\B|,W
+\lambda,Q)|\\
\le& C W_-^{\gamma+\frac12}[|\log W_-|+1] |Q|\max_{Q^{(2)}}|B(x)|.\end{array}
\end{equation}
Both (\ref{Eqn5.19}) and (\ref{Eqn5.20}) hold uniformly in
$\mu\ge 0$. \end{Lemma} \begin{proof} By Theorem~\ref{Count}
\begin{equation}\label{Eqn5.21} \lim_{\hbar\to 0}\sup_{\mu\hbar\le
C}\frac{\hbar^3}{\mu\hbar+1}| N(\mathbb{P}_W+\lambda,Q) -
\mathfrak{B}(\mu\hbar |\B|,W+\lambda,Q)| = 0.
\end{equation}
From
Lemma~\ref{Thm5.2}
$$
\frac{\hbar^3}{\mu\hbar+1}N(\mathbb{P}_W+\lambda,Q)\lsim
|\log\lambda^{-1}|(W+\lambda)_-^\frac12 |Q|\max_{Q^{(2)}}|B(x)| $$
and
$$\begin{array}{l}
\frac{\hbar^3}{\mu\hbar+1}\hbar^{-3}\mathfrak{B}_0(\mu\hbar|B|,W+\lambda,Q
)\\
\ \ \ \le \int_Q|B|[(W+\lambda)_-^\frac12+2\sum_{k\ge
1}(2k\mu\hbar|B|+W+\lambda)_-^\frac12]d\mathbf{x}\\
\ \ \ \lsim
\int_Q(W+\lambda)_-^\frac12[|B|+(\mu\hbar)^{-1}(W+\lambda)_-]d\mathbf{x}.
\end{array}
$$
Therefore,
$$\begin{array}{l}
\sup_{\mu\hbar\ge C}\frac{\hbar^3}{\mu\hbar+1}| N(\mathbb{P}_W+\lambda,Q)
- \hbar^{-3}\mathfrak{B}(\mu\hbar |\B|,W+\lambda,Q)|\\ \ \ \ \lsim
|\log\lambda^{-1}|(W+\lambda)_-^\frac12
|Q|\max_{Q^{(2)}}|B(x)|+\frac{1}{C}|Q|(W+\lambda)_-^\frac32
\end{array}
$$
and (\ref{Eqn5.19}) follows since $C$ is arbitrary.
The proof of (\ref{Eqn5.20}) follows in a similar manner.
\end{proof}

\begin{Theorem}\label{Thm5.6}
Let $ \B(\mathbf{x})=
(0,0,B(x))$. Suppose that
\begin{enumerate}
\item
$B$ is continuous,
\item
$W_- \in L^{\infty}(\Omega)$,
\item
for some $p>1, W_-^2, \hat{b}_pW_- \in L^1(\Omega)$, where $
\hat{b}_p $ is defined in (\ref{bphat}).
\end{enumerate}
Then, for all $\gamma \ge 0$ and $\lambda >0 $,
\begin{equation}\label{Eqn5.22} \lim_{\hbar\to 0}
\frac{\hbar^3}{\mu\hbar +1}\left\{M_{\gamma}(\mathbb{P}_{W}+\lambda,\Omega)-
\hbar^{-3}\mathfrak{B}_{\gamma}(\mu\hbar|\B|,W+\lambda,\Omega)\right\}=0.
\end{equation}
uniformly for $\mu\ge 0$. If $\gamma>1/2 $, (\ref{Eqn5.22}) holds for
$\lambda =0$.
\end{Theorem}
\begin{proof}
We prove the result for $\gamma =0$, the proof for $\gamma >0$
being similar. The last assertion in the theorem, concerning
$\gamma >1/2$, is proved in \cite{Sobolev1998}.
Let $W_0$, $Q_k$, $k=1,2,\dots,k $ be as in \S2 ($A_1$)-($A_3$).
As in \S6 of
Sobolev~\cite{Sobolev1998}, for $\eta >0$, we partition each $Q_k$
into a finite number of cubes $Q_{jk}$, $j=1,2,\dots$, such that
for each pair $k,j$
\begin{equation}\label{Eqn5.24}
|B(x)-B(y)|\le \eta,\q x,y\in Q_{k,j}.
\end{equation}
Let $I_+\equiv I_+(\eta):=\{(k,j):\max_{Q_{k,j}}|B(x)|\ge
2\eta\}$ and $I_-\equiv I_-(\eta)$ the complementary set. In view
of (\ref{Eqn5.24}) we have that $|B(x)|\ge \eta$ for $x\in
Q_{k,j}$, $(k,j)\in I_+$.
Since $\mathbb{P}_{W_0}(\Omega)\le \oplus_{k=1}^K\mathbb{P}_{W_0}(Q_k)$,
$$ N(\mathbb{P}_{W_0}+\lambda,\Omega)\ge
\sum_{k=1}^KN(\mathbb{P}_{W_0}+\lambda,Q_k)
$$
and from Corollary~\ref{Cor5.4} and Lemma~\ref{Lem5.5}
$$
\begin{array}{l} \liminf_{\hbar\to
0}\frac{\hbar^3}{\mu\hbar +1}\left\{N(\mathbb{P}_{W_0}+\lambda,\Omega)
-\hbar^{-3}\mathfrak{B}_0(\mu\hbar|\B|,W_0+\lambda,\Omega)\right \}\\ \ge
-C\eta\sum_{(k,j)\in I_-}\int_{Q_{k,j}}(W_0+\lambda)_-^\frac12 dx
\gsim -\eta. \end{array} $$ Since $\eta$ is arbitrary, then
\begin{equation}\label{Eqn5.25} \liminf_{\hbar\to
0}\frac{\hbar^3}{\mu\hbar+1}\left\{ N(\mathbb{P}_{W_0}+\lambda,\Omega)
-\hbar^{-3}\mathfrak{B}_0(\mu\hbar|\B|,W_0+\lambda,\Omega)\right\}\ge 0
\end{equation}
for each $\lambda >0$. For the reverse inequality we proceed as in the
proof of (\ref{Eqn3.10}). Let the interior of each $Q_k$ be denoted by
$int(Q_k)$. Set
\begin{equation}\label{Eqn5.3A}\begin{array}{l}
S:=\Omega \setminus \cup_{k=1}^K int(Q_k),\q  {S}_{\rho r} :=\{x\in\Omega:
dist(x,S)<\rho r\},\\
 Q(\rho r):=\{x\in\Omega:dist(x,Q)<\rho r\}\end{array}
\end{equation}
for some $\rho\in (0,1)$.
 Construct a partition of unity  $\{\psi_k\}_{k=-1}^K$ subordinate
 to the covering $\cup_{k=1}^K int (Q_k)\cup S_{\rho r}$ of $\Omega$:
\begin{equation}\label{Eqn5.4A}\begin{array}{rl}
(i) & \psi_{-1}\in C^\infty(\Omega),\ \ \psi_0\in\Con(S_{\rho r}\cap
Q_{\rho r}),\ \
\psi_k\in \Con(int (Q_k)),\ k=1,\dots,K;\\
(ii) & \sum_{k=-1}^K\psi_k^2 \equiv 1, \forall x;\\
(iii) & \psi_{-1}(x)=0,\ x\in Q,\ \ \psi_{-1}(x)=1,\ x\notin Q(\rho r);\\
(iv) & \sum_{k=-1}^K |\grad \psi_k(x)|^2 \le C(\rho r)^{-2}\chi_{S_{\rho
r}\cap Q(\rho r)}\end{array}
\end{equation}
for some $C>0$ where
$\chi_{\rho r}$ is the characteristic function for
$S_{\rho r}\cap Q(\rho r)$. It follows that
\begin{equation}\label{Eqn5.26}
|S_{\rho r}\cap Q|\lsim \rho |Q|,\q |Q(\rho r)\setminus Q|\lsim \rho r,
\end{equation}
and for every $f\in [\Con(\Omega)]^2$
$$
\sum_{k=-1}^K(\mathbb{P}_0\psi_kf,\psi_kf) \le (\mathbb{P}_0 f,f)
+C\hbar^2(\rho r)^{-2}(\chi_{S_{\rho r}\cap Q(\rho r)}f,f).
$$
Let $r=A\hbar$ and $\lambda-C(A\rho)^{-2} \ge0$. Then,
\begin{equation}\label{Eqn5.6A}\begin{array}{rl}
((\mathbb{P}_{W_0}+\lambda)f,f)\ge & \sum_{k=-1}^K([\mathbb{P}_{W_0}
+\lambda-C\hbar^2(\rho r)^{-2}\chi_{\rho r}]\psi_k
f,\psi_k f)\\ \ge & \sum_{k=0}^K([\mathbb{P}_{W_0} +\lambda-C\hbar^2(\rho
r)^{-2}\chi_{\rho r}]\psi_k f,\psi_k f) \end{array}
\end{equation}
since $\psi_{-1}\equiv 0$ in $Q$,and $W_0=0$ for $x\notin Q$. This implies that
\begin{equation}\label{Eqn5.27}\begin{array}{rl}
N(\mathbb{P}_{W_0}+\lambda,\Omega)\le &
N(\mathbb{P}_{W_0}+\lambda-C\hbar^2(\rho r)^{-2},S_{\rho r}\cap Q(\rho
r))\\ &+\sum_{k=1}^KN(\mathbb{P}_{W_0}+\lambda-C\hbar^2(\rho r)^{-2},Q_k).
\end{array} \end{equation}
From Corollary~\ref{Cor5.4}
and Lemma~\ref{Lem5.5}
\begin{equation}\label{Eqn5.28}\begin{array}{rl}
\limsup_{\hbar\to 0}\frac{\hbar^3}{\mu\hbar +1}
\sum_{k=1}^K&\{N(\mathbb{P}_{W_0}+\lambda-\frac{C\hbar^2}{(\rho
r)^{2}},Q_k)\\
&\ \ -
\hbar^{-3}\mathfrak{B}_0(\mu\hbar|\B|,W_0+\lambda-\frac{C\hbar^2}{(\rho
r)^2},Q_k)\}\\ \le & C\eta\sum_{(k,j)\in
I_-}\int_{Q_{k,j}}(W_0+\lambda)_-^\frac12 d\mathbf{x} \lsim \eta.
\end{array}
\end{equation}
and from (\ref{Eqn2.10}),
\begin{equation}\label{Eqn5.29}
\begin{array}{l} \frac{1}{\mu\hbar
+1}|\mathfrak{B}_0(\mu\hbar|\B|,W_0+\lambda-\frac{C\hbar^2}{(\rho
r)^2},Q_k)-\mathfrak{B}_0(\mu\hbar|\B|,W_0+\lambda,Q_k)|\\ \lsim
\frac{1}{A\rho}+\frac{1}{(A\rho)^3}.
\end{array}
\end{equation}
From (\ref{Eqn5.2}) and (\ref{Eqn5.26})
\begin{equation}\label{Eqn5.30}\begin{array}{l} \mu^{-1}\hbar^2
N(\mathbb{P}_{W_0}+\lambda-C\hbar^2(\rho r)^{-2},S_{\rho r}\cap Q(\rho
r))\\ \lsim e^{1/(A\rho)^2}\frac{1}{(A\rho)^3}\rho |Q|.
\end{array}
\end{equation}
It now follows from (\ref{Eqn5.27})-(\ref{Eqn5.30}), on allowing $\hbar \rightarrow 0,
A\rightarrow \infty, \eta \rightarrow 0, \rho \rightarrow 0 $, in that order,
that
$$
 \limsup_{\hbar\to
0}\frac{\hbar^3}{\mu\hbar +1}\{N(\mathbb{P}_{W_0}+\lambda,\Omega)
-\hbar^{-3}\mathfrak{B}_0(\mu\hbar|\B|,W_0+\lambda,\Omega)\}\le 0
$$
which together with (\ref{Eqn5.25}) implies
\begin{equation}\label{piecewisecase}
 \lim_{\hbar\to
0}\frac{\hbar^3}{\mu\hbar +1}\{N(\mathbb{P}_{W_0}+\lambda,\Omega)
-\hbar^{-3}\mathfrak{B}_0(\mu\hbar|\B|,W_0+\lambda,\Omega)\}= 0.
\end{equation}
Similarly, we have for any $\gamma>0$ and $\lambda > 0$
\begin{equation}\label{N106}
 \lim_{\hbar\to
0}\frac{\hbar^3}{\mu\hbar +1}\{M_{\gamma}(\mathbb{P}_{W_0}+\lambda,\Omega)
-\hbar^{-3}\mathfrak{B}_{\gamma}(\mu\hbar|\B|,W_0+\lambda,\Omega)\}= 0.
\end{equation}
The proof for general $W$ is similar to the corresponding part of
the proof of Theorem~\ref{Count2} in \S3, using the Weyl inequalities
in the same way as Sobolev does in \cite{Sobolev1998}, but, instead of (\ref{N46}) we now use
the inequality derived by Shen in Lemmas 5.1, 5.3, 5.4 and (5.14) of \cite{Shen1999},
namely,
\begin{equation}\label{N107}
\hbar^3 N(\pw+\lambda, \mathbb{R}^3) \lsim \frac{1}{\sqrt \lambda
} \big \{\int_{ \mathbb{R}^3} |W_-( \mathbf{x})|^2 d\mathbf{x}
+\mu \hbar \int_{ \mathbb{R}^3}\hat{b}_p( \mathbf{x})
W_-( \mathbf{x}) d\mathbf{x}\big\}.
\end{equation}
\end{proof}

The following analogue of Theorem~\ref{New3} is also readily
proved.

\begin{Theorem}\label{NewThm}
Let $\B(\mathbf{x})= (0,0,B(x))$. Suppose that
\begin{itemize}
\item [(i)] $W$, $\B$ are continuous,
\item [(ii)] $|W(\mathbf{x})|\to 0$ uniformly as $|\mathbf{x}|\to
\infty $ in $\Omega$.
\end{itemize}
Then, (\ref{Eqn5.22}) holds for all $\gamma\in [0,1)$ and $\lambda
>0$.
\end{Theorem}

Finally in this section we give the result promised after
Lemma~\ref{Thm5.2}.

\begin{Proposition}\label{Prop5}
Let $\Omega = \Omega^{(2)}\times\R$, where $\Omega^{(2)}$ is of
finite measure in $\R^2$, and suppose that
\begin{itemize}
\item [(i)] $\B=(0,0,B(x))$ and $|\B|\in L^\infty(\Omega^{(2)})$;
\item [(ii)] $\inf_{\Omega}W(\mathbf{x})\ge -c\chi_{_{(-R,R)}}(x_3)$,
$\mathbf{x}=(x,x_3)$, for some positive numbers $c$, $R$.
\end{itemize}
Then for $\lambda >0$
\begin{equation}\label{A}
N(\mathbb{P}_W+\lambda,\Omega) \lsim \hbar^{-3}(\mu\hbar
+1)\sqrt{c}|\log \lambda^{-1}|R\int_{\Omega^{(2)}}(|B(x)|+c)dx
\end{equation}
and, for any $\gamma >0$,
\begin{equation}\label{B}
M_\gamma (\mathbb{P}_W,\Omega)\lsim \hbar^{-3}(\mu\hbar
+1)c^{\gamma+\frac12}(|\log c|+1)R\int_{\Omega^{(2)}}(|B(x)|+c)dx.
\end{equation}
\end{Proposition}
\begin{proof}
In the notation of the proof of Lemma~\ref{Thm5.2},
$$
\mathbb{P}_W\ge \mathbb{P}_0^{(2)} + (-\hbar^2\partial_3^2-\tilde
W)
$$
where $\tilde W(\mathbf{x})= \tilde W(x_3)=c\chi_{_{(-R,R)}}(x_3)$.
The expression $-\hbar^2\partial_3^2-\tilde W$ has a self-adjoint
realization in $L^2(\R)$ with essential spectrum $[0,\infty)$ (see
\cite{Raikov1999}, Proposition~2.1), and it is readily shown that there are
negative eigenvalues at the solutions of the equation
\begin{equation}\label{B11}
\tan(\frac{2R}{\hbar}\sqrt{c-\lambda}) =
4\frac{\sqrt{\lambda(c-\lambda)}}{c-2\lambda},\q \lambda\in (0,c)
\end{equation}
corresponding to the eigenfunctions
\begin{equation}\label{B22}\begin{array}{l}
\varphi(x_3,\lambda)=\\
\left\{\begin{array}{l}
\frac{e^{2R\sqrt{\lambda}/\hbar}}{2i\theta}
\big [(1+i\theta)e^{(1+i\theta)
\frac{R\sqrt{\lambda}}{\hbar}-i\theta\frac{x_3\sqrt{\lambda}}{\hbar}}
-(1-i\theta)e^{(1-i\theta)\frac{R\sqrt{\lambda}}{\hbar}+i\theta\frac{x_3\sqrt{\lambda}}{\hbar}}
\big ],
\ \ |x_3|\le R,\\
e^{-\frac{\sqrt{\lambda}}{\hbar}x_3},\ \ \ x_3>R,\\
e^{\frac{\sqrt{\lambda}}{\hbar}x_3},\ \ \
x_3<-R,\end{array}\right.\end{array}
\end{equation}
where $\theta:=\sqrt{\frac{c}{\lambda}-1}$.
If these negative eigenvalues are denoted by $-\epsilon_m$,
$m=1,\dots,M$, then $\epsilon_m\le c$ and
\begin{equation}\label{B33}
M\le\#\{n:c-\frac{(n\pi\hbar)^2}{4R^2}\ge 0\} =
\left [2\frac{R\sqrt{c}}{\pi\hbar}\right ].
\end{equation}
We now proceed in a similar way to the proof of
Lemma~\ref{Thm5.2}. We have that for any $\epsilon \in (0,1)$
\begin{equation}\label{B44}\begin{array}{rl}
N(\mathbb{P}_W+\lambda,\Omega)\le & \sum_{m=1}^M
N(\mathbb{P}_0^{(2)}+\lambda-\epsilon_m,\Omega^{(2)})\\
\le &
2\sum_{m=1}^M N(\epsilon
H_0^{(2)}-\epsilon\mu\hbar|B|+\lambda-\epsilon_m,\Omega^{(2)}).\end{array}
\end{equation}
Now substitute
$$
\mathcal{A}=\epsilon H_0^{(2)}+\lambda,\ \
\mathcal{B}=-\epsilon\hbar^2\Delta +\lambda,\ \ V=(\epsilon\mu\hbar
|B|+\epsilon_m)\chi_{_{\Omega^{(2)}}}
$$
in Theorem~2.4 of \cite{RozenblumSolomyak1997}. This gives, as in
(\ref{Eqn5.6}),
$$
N(\mathcal{A}-V)\le \frac{1}{2\pi\hbar^2\epsilon
g(1)}\int_{\Omega^{(2)}} V(x)dx \int_{\phi(x)}^\infty
e^{-\lambda t}(t-\phi(x))\frac{dt}{t^2}
$$
where $\phi(x)=k/V(x)$ and $g(1)$ satisfies (\ref{Eqn5.7}). Choose
$$
\epsilon =\frac{1}{\mu\hbar +1},\ \ k=\max_{\Omega^{(2)}}|B(x)|+c.
$$
Then $\phi(x)\ge 1$ and the result follows as before.
\end{proof}

\section{The Dirac operator: magnetic fields with constant direction.}
\begin{Theorem}\label{Thm6}
Suppose that
\begin{enumerate}
\item
$B$ is continuous,
\item
$V_- \in L^{\infty}(\Omega)$,
\item
for some $p>1$, $V_-^4$, $\hat{b}_p V_-^2 \in L^1(\Omega)$,
\item
$|\{ \mathbf{x} \in \Omega: |V( \mathbf{x})| >2 \}| =0$.
\end{enumerate}
Then, for all $\gamma \ge 0 $,
$$\begin{array}{rl}
\lim_{\lambda \rightarrow 0^+}\lim_{\hbar \rightarrow 0}&\big(\frac{\hbar^3}{\mu \hbar
+1}\big)\big \{\frac12[M_{\gamma} (\dv, \Omega, I(\lambda))+
M_{\gamma} ( \mathbb{D}_{-V}, \Omega, I(\lambda))]\\
& -\hbar^{-3}\mathfrak{B}_{\gamma}(\mu \hbar |B|, \lambda_1,
\Omega) \big \} =0,\end{array}
$$
where $\lambda_1 = 2|V|+V^2$ and $I(\lambda)=(-\sqrt{1-\lambda},\sqrt{1-\lambda})$.
\end{Theorem}
\begin{proof}
The proof follows from Theorem~\ref{Thm5.6} just as Theorem~\ref{Thm4} follows from
Theorem~\ref{Count2}. As for (\ref{Eqn4.3}) and (\ref{Eqn4.8}), we show that (in the notation of
($A_1$)-($A_3$) in \S2)
\begin{equation}\label{N114}\begin{array}{rl}
\liminf_{\hbar \rightarrow 0}&(\frac{\hbar^3}{\mu \hbar
+1})\big\{(1/2)\big[N((\mathbb{D}_{V_0},\Omega,I(\lambda)) +
N((\mathbb{D}_{-V_0},\Omega,I(\lambda))\big]\nonumber\\
&\ \ \ \ \ \ \   -\hbar^{-3}\big[ \mathfrak{B} (\mu \hbar|B|,-\lambda_1^0, \Omega) +
\mathfrak{B} (\mu \hbar|B|,-\lambda_{-1}^0, \Omega)\big] \big
\} \ge
0\end{array}
\end{equation}
where $\lambda_{\pm 1}^0(\lambda)=(\sqrt{1-\lambda}\pm |V_0|)^2-1$, and
\begin{equation}\label{N115}\begin{array}{rl}
\limsup_{\hbar \rightarrow 0}&(\frac{\hbar^3}{\mu \hbar
+1})\big\{(1/2)\big[N((\mathbb{D}_{V_0},\Omega,I(\lambda)) +
N((\mathbb{D}_{-V_0},\Omega,I(\lambda))\big]\nonumber\\
& \ \ \ \ \ \ \ -\hbar^{-3}
\big[ \mathfrak{B} (\mu \hbar|B|,-\lambda_1^0, \Omega) -
\mathfrak{B} (\mu \hbar|B|,-\lambda_{-1}^0, \Omega)\big] \big\} \le
0\end{array}
\end{equation}
The rest of the proof is similar to that of Theorem~\ref{Thm4}.
\end{proof}
\section{Appendix}
\subsection{Approximation of gauges in cubes}
Let $\mathbf{x}_{Q_k}=(x_{Q_k}^1,x_{Q_k}^2,x_{Q_k}^3)$ be the center of the cube
$Q_k$. Since
 div$(\B)=0$, we may choose the gauge $\sa=(a_1,a_2,0)$
$$
\begin{array}{rl}
a_1(\mathbf{x}):=&-\frac12\int_{x_{Q_k}^2}^{x_2}B_3(x_1,t,x_{Q_k}^3)dt +
\int_{x_{Q_k}^3}^{x_3}B_2(x_1,x_2,t)dt,\\
a_2(\mathbf{x}):=&\frac12\int_{x_{Q_k}^1}^{x_1}B_3(t,x_2,x_{Q_k}^3)dt -
\int_{x_{Q_k}^3}^{x_3}B_1(x_1,x_2,t)dt
\end{array}
$$
-- see the proof of Lemma~3.1 of Sobolev~\cite{Sobolev1998}. Let
$\B^o(\mathbf{x}):=\B(\mathbf{x}_{Q_k})$, $\mathbf{x}\in Q_k$ and define the piecewise linear
vector potential  $\AAAA$ by
\begin{eqnarray*}
\AAA_1(\mathbf{x})&:=&-\frac12
B_3(\mathbf{x}_{Q_k})(x_2-x_{Q_k}^2)+B_2(\mathbf{x}_{Q_k})(x_3-x_{Q_k}^3),\\
\AAA_2(\mathbf{x})&:=&\frac12
B_3(\mathbf{x}_{Q_k})(x_1-x_{Q_k}^1)-B_1(\mathbf{x}_{Q_k})(x_3-x_{Q_k}^3),\\ \quad
\AAA_3:=&0.
\end{eqnarray*}
for $ \mathbf{x}\in Q_k$. Then $ \grad \times  \AAAA = \B^o$ in
$Q_k$ and (11) is satisfied.

\bibliographystyle{amsalpha}

\end{document}